\documentclass[]{aastex7}
\usepackage{ulem}
\usepackage{mhchem}
\usepackage{bm}

\newcommand\subs[1]{\textsubscript{#1}}
\newcommand\sups[1]{\textsuperscript{#1}}
\newcommand\rh[1]{\textcolor{black}{{\textit{r}\subs{\textit{H}}}#1}}

\newcommand\trot[1]{\textcolor{black}{{\textit{T}\subs{rot}}#1}}

\newcommand\kms[1]{\textcolor{black}{{km\,s$^{-1}$}#1}}
\newcommand\ps[1]{\textcolor{black}{{s$^{-1}$}#1}}

\newcommand\um[1]{\textcolor{black}{{$\mu$m}#1}}

\definecolor{gold}{rgb}{0.64,0.54,0.29}

\received{}
\revised{}
\accepted{}
\submitjournal{The Astrophysical Journal}
\makeatletter
\renewcommand{\frontmatter@title@above}{}
\makeatother

\shorttitle{JWST Studies of C/2022 E3}
\shortauthors{Milam et al.}

\begin{document}


\title{JWST Spatial-Spectral Mapping of Green Comet C/2022 E3 (ZTF)}

\correspondingauthor{Stefanie N. Milam}
\email{stefanie.n.milam@nasa.gov}

\author[0000-0001-7694-4129]{Stefanie N. Milam}
\affiliation{Solar System Exploration Division, NASA Goddard Space Flight Center, 8800 Greenbelt Rd, Greenbelt, MD 20771, USA}
\email{stefanie.n.milam@nasa.gov}

\author[0000-0002-6006-9574]{Nathan X. Roth}
\affiliation{Solar System Exploration Division, NASA Goddard Space Flight Center, 8800 Greenbelt Rd, Greenbelt, MD 20771, USA}
\affiliation{Department of Physics, American University, 4400 Massachusetts Ave NW, Washington, DC 20016, USA}
\email{nathaniel.x.roth@nasa.gov}

\author[0000-0002-2662-5776]{Geronimo L. Villanueva}
\affiliation{Solar System Exploration Division, NASA Goddard Space Flight Center, 8800 Greenbelt Rd, Greenbelt, MD 20771, USA}
\email{geronimo.l.villanueva@nasa.gov}

\author{Dominique Bockelée-Morvan}
\affiliation{LIRA, Observatoire de Paris, Université PSL, CNRS, Sorbonne Université, Université Paris Cité, 5 place Jules Janssen, 92195 Meudon, France}
\email{dominique.bockelee@obspm.fr}

\author{Jacques Crovisier}
\affiliation{LIRA, Observatoire de Paris, Université PSL, CNRS, Sorbonne Université, Université Paris Cité, 5 place Jules Janssen, 92195 Meudon, France}
\email{jacques.crovisier@obspm.fr}

\author[0000-0001-6397-9082]{David E. Harker}
\affiliation{Department of Astronomy and Astrophysics, University of California, San Diego, 9500 Gilman Drive, MC 0424, La Jolla, CA 92093-0424, USA}
\email{dharker@ucsd.edu}

\author[0000-0002-6702-7676]{Michael S. P. Kelley}
\affiliation{Department of Astronomy, University of Maryland, College Park, MD 20742-0001, USA}
\email{msk@astro.umd.edu}

\author[0000-0001-6192-3181]{Kiernan D. Foster}
\affiliation{Department of Chemistry, University of Virginia, Charlottesville, VA 22904, USA}
\email{kfoster@virginia.edu}

\author[0000-0001-9665-8429]{Ian Wong}
\affiliation{Space Telescope Science Institute, 3700 San Martin Drive, Baltimore, MD, USA}
\email{iwong@stsci.edu}

\author[0000-0003-0774-884X]{Davide Farnocchia}
\affiliation{Jet Propulsion Laboratory, California Institute of Technology, Pasadena, CA, USA}
\email{davide.farnocchia@jpl.nasa.gov}

\author[0000-0001-7895-8209]{Marco Micheli}
\affiliation{ESA NEO Coordination Centre, Planetary Defence Office, European Space Agency, Frascati, Italy}
\email{marco.bs.it@gmail.com}

\author[0000-0001-8751-3463]{Heidi B. Hammel}
\affiliation{Association of Universities for Research in Astronomy, Suite 1475, 1331 Pennsylvania Avenue NW, Washington DC 20004, USA}
\email{hbhammel@aura-astronomy.org}




\begin{abstract}

We report a survey of molecular emission from cometary volatiles using the James Webb Space Telescope (JWST) toward Oort cloud comet C/2022 E3 (ZTF) carried out on UT 2023 February 28 and March 1 at a heliocentric distance (\rh{}) of 1.33 au. These measurements of \ce{H2O}, HCN, \ce{CH3OH}, \ce{C2H6}, \ce{CH4}, CO, \ce{CO2}, \ce{^13CO2}, and OCS sampled post-perihelion molecular chemistry in C/2022 E3. A suite of near$-$mid-infrared OH$^*$ (prompt emission) transitions were also detected. This work presents nucleus-centered spectra for all detected species, spatial-spectral maps of column density and rotational temperature as a function of distance from the nucleus for all except \ce{C2H6}, HCN, and OH*, and maps of co-measured continuum. The spatial distributions of both quantities were anisotropic for all mapped molecules; however, \ce{H2O} showed distributions distinct from the remaining species. Coma-averaged values of the ortho-to-para ratio (OPR) for \ce{H2O} and the \ce{^12CO2}/\ce{^13CO2} ratio derived from these maps were consistent with the statistical equilibrium value of 3 and the terrestrial value of 89, respectively. The modeled mass fraction of the sub-micron dust grains is dominated by amorphous carbon ($\sim56\%$) followed by amorphous Mg:Fe pyroxene ($\sim28\%$), Mg-rich crystalline olivine ($\sim10\%$), and amorphous Mg:Fe olivine ($\sim5\%$) with a crystalline mass fraction of $0.2385\pm0.0008$. We compare the volatile and dust composition of C/2022 E3 (ZTF) against comets measured to date, including those surveyed by JWST.
\end{abstract}



\section{Introduction} \label{sec:intro}
Comets contain primitive material preserved from the birth of the solar system. Assembled at the time of planet formation, they likely incorporated ices from the interstellar medium, as well as material mixed over wide ranges in the protoplanetary disk \citep[e.g.,][]{Bockelee2004,Zolensky2006,MummaCharnley}. Understanding how cometary nuclei were assembled is key to unraveling what they can tell us about the birth and evolution of the early solar system. 

Remote sensing studies of cometary atmospheres typically produce several compositional metrics, including: (1) Mixing ratios (relative abundances) of coma volatiles and/or (2) Spatial maps of molecular column density and/or temperature. Previous work has revealed significant compositional diversity among the comet population \citep{DelloRusso2016a,Lippi2021,Bockelee2017,Biver2024b}, and recent model-data synergies have explored the ice-phase protoplanetary disk chemistry which may explain the molecular abundances measured in solar system comets \citep[e.g.,][]{Willacy2022,Willacy2025}. On the other hand, spatial studies of the distribution of molecular column density and rotational temperature at near-infrared wavelengths have been previously limited to 1D spatial profiles produced by ground-based long-slit spectroscopy \citep{Villanueva2011a,DelloRusso2016b,Bonev2013,Bonev2014,Faggi2021}. Such studies have indicated distinct spatial associations among coma volatiles, which provide insights into how ices are associated or segregated in the nucleus, as well as revealing coma heating and cooling mechanisms through studies of H$_2$O rotational temperature profiles. 

The sensitivity and spatial-spectral mapping capabilities of the NIRSpec and MIRI \citep{Rigby2023,Gardner2023} IFUs are now enabling 3D spatial-spectral mapping of cometary atmospheres which was previously only available with millimeter/sub-millimeter wavelength interferometers such as the Atacama Large Millimeter Array or the Northern Extended Millimeter Array \citep[e.g.,][]{Boissier2007,Cordiner2023}. Multi-species JWST spatial-spectral studies of comets 29P/Schwassmann-Wachmann 1 \citep{Faggi2024}, C/2017 K2 \citep[PanSTARRS;][]{Woodward2025}, and 3I/ATLAS \citep{Roth2026c} demonstrated significant differences in the spatial distributions of column density and rotational temperatures among species in those comets. 

Here we report the complete JWST near- to mid-infrared spectrum of C/2022 E3 (ZTF), obtained on 2023 February 28 and March 1. Three NIRSpec IFU gratings (G140H/F100LP, G235H/F170LP, G395H/F290LP) and all MIRI MRS bands/channels revealed the molecular abundances of a dynamically old Oort cloud comet \citep{Nakano2023}. For species with a sufficiently high signal-to-noise ratio, spatial-spectral maps of column density and rotational temperature were obtained. Our analysis includes the 1.4, 1.9, and 6 \um{} \ce{H2O} bands, CO, \ce{CO2}, \ce{^13CO2}, \ce{CH4}, \ce{C2H6}, HCN, OH$^*$ \citep[prompt emission, the vibrationally and rotationally excited photolysis product of \ce{H2O};][]{Bonev2006}, and CN, building upon studies of \ce{H2O} (specifically the 2.6, 2.9, and 4.5 \um{} bands), \ce{CH3OH}, and OCS in these same JWST data by \cite{Foster2026}.

\section{Observations and Data Reduction} \label{sec:obs}
Comet C/2022 E3 (ZTF; hereafter E3) reached perihelion ($q=1.11$ au) on 2023 January 12. We conducted post-perihelion observations of E3 using the JWST MIRI MRS and NIRSpec IFUs. E3's \rh{} ranged from $1.33-1.34$ au, its distance from the telescope ($\Delta_{\mathrm{JWST}}$) from $0.91-0.94$ au, and the solar phase angle ($\phi$) from $48.2\degr-47.7\degr$. Observations used a representative resolving power $\lambda/\Delta\lambda\sim2700$ for NIRSpec, whereas the MIRI resolving power ranges from $\sim3500$ in Channel 1 ($4.90-7.65$ \um{}) to $\sim1700$ in Chanel 4 ($17.7-27.9$ \um{}). The NIRSpec G140H/F100LP (spanning $0.97–1.89$ \um{}) samples emission from CN and \ce{H2O}, NIRSpec G235H/F170LP ($1.66–3.17$ \um{}) includes the \ce{H2O} $\nu_1+\nu_3$ band near 2.7 \um{}, and NIRSpec G395H/F290LP covers CO ($v=1-0$), \ce{CO2} ($\nu_3$, $\nu_1+\nu_3-\nu_1$), \ce{^13CO2} ($\nu_3$), OCS ($\nu_3$), \ce{CH3OH} ($\nu_2, \nu_3, \nu_9$), \ce{CH4} ($\nu_3$), \ce{C2H6} ($\nu_7$), HCN ($\nu_1)$, \ce{C2H2} ($\nu_2$), \ce{NH3} ($\nu_1$), \ce{NH2}, OH*, and multiple \ce{H2O} hot bands near 2.9 \um{} and 4.5 \um{}. MIRI MRS includes \ce{H2O} ($\nu_6$), OH$^*$, and the \ce{CO2} hot bands near 15 \um{}. Background exposures with identical circumstances to comet frames (exposure time, grating settings) were conducted offset from the comet position by $300''$. Exposures were processed using the JWST Pipeline version 1.16.0 with CRDS jwst\_1298.pmap context files and aligned onto a common spatial-spectral axis using the Drizzle algorithm \citep{Law2023}. An observing log is provided in Table~\ref{tab:obslog}.

\begin{deluxetable*}{cccccccc}
\tablenum{1}
\tablecaption{Observing Log\label{tab:obslog}}
\tablewidth{0pt}
\tablehead{
\colhead{Date} & \colhead{Setting} & \colhead{UT Time} & \colhead{\textit{t}\subs{int}} &
\colhead{\textit{r}\subs{H}} & \colhead{$\Delta_{\mathrm{JWST}}$} & \colhead{$\phi_\mathrm{STO}$} & \colhead{$\lambda$} \\
\colhead{(2023)} & \colhead{} & \colhead{} & \colhead{(s)} & \colhead{(au)} & 
\colhead{(au)} & \colhead{($\degr$)} & \colhead{($\mu$m)}
}
\startdata
28 Feb & MIRI MRS & 06:38 & 222 & 1.33 & 0.91 & 48.2 & $4.90-27.90$ \\
01 Mar & G395H/F290LP & 04:33 & 1517 & 1.34 & 0.94 & 47.8 & $2.87-5.14$ \\
       & G140H/F100LP & 07:07 & 933 & 1.34 & 0.94 & 47.8 & $0.97-1.89$ \\
       & G235H/F170LP & 07:37 & 816 & 1.34 & 0.94 & 47.8 & $1.66-3.17$ \\
\enddata
\tablecomments{UT times are given at the start of each exposure. \textit{t}\subs{int} is the total on-source integration time. \textit{r}\subs{H}, $\Delta_{\mathrm{JWST}}$, and $\phi_\mathrm{STO}$, and $\psi_{\sun}$ are the heliocentric distance, JWST-centric distance,
and phase angle (Sun--Comet--Earth), respectively, of C/2022 E3 at the time of observations. }
\end{deluxetable*}

We generated spatial-spectral maps of molecular column density ($N$, m$^{-2}$) and rotational temperature (\trot{}, K) by performing a spaxel-by-spaxel analysis of the MIRI MRS and NIRSpec IFU data cubes following the methods of \cite{Roth2026c, Woodward2025}. Each modeled quantity was retrieved using molecular emission models in the NASA Planetary Spectrum Generator \citep[PSG;][]{Villanueva2018}. All spectra were analyzed using the PSG following its significant update in 2026 March\footnote{https://psg.gsfc.nasa.gov/about.php}. We extracted and modeled spectra from each $0\farcs1$ spaxel using the \texttt{jwstComet} package \citep{Roth2026b}, which provides for flexible spectral extraction from JWST IFU data cubes using functions from the \texttt{astropy} and \texttt{photutils} libraries, followed by automated interfacing with the NASA PSG API. The data cubes were converted from units of MJy sr$^{-1}$ to Jy pixel$^{-1}$, then spectra were extracted from the data cubes using the \texttt{photutils} RectangularAperture function in combination with the aperture\_photometry function (method = `center'). Fluxes and $1\sigma$ instrumental noise on a per-spaxel basis were derived from the \texttt{SCI} and \texttt{ERR} extensions of the FITS files.

Contributions from gaseous and continuum emission were identified by comparing the spectra with expected spectral line positions for each species from quantum mechanical models of fluorescent emission generated with the PSG. Each spectrum was baseline subtracted using low-order (second to fourth) polynomial baselines. This baseline accounts for continuum emission from the dust and nucleus, as well as scattered sunlight and instrumental artifacts. We chose the lowest possible polynomial order that could account for the spectral shape while avoiding higher order polynomials to prevent introducing spurious features into the spectra. The baselines were fit simultaneously with the molecular emission models using the Optimal Estimation Method implemented in the PSG. Techniques employing simultaneous fitting of the continuum baseline and molecular emission models have been applied to decades of cometary infrared spectroscopy studies \citep[e.g.,][]{DiSanti2003,Villanueva2011a,Bonev2014,DiSanti2017,Roth2018,Faggi2019,Ejeta2024}. Fitting the baselines and emission models simultaneously ensured that uncertainties on the baseline fit were propagated into uncertainties on each retrieved quantity (i.e., $N$ or \trot{}). These fits included a correction for opacity effects on a spaxel-by-spaxel basis \citep[see ][ for further details]{Roth2026c,Villanueva2025,Roth2023}. However, it is worth nothing that the solar phase angle during our observations was $\phi\sim48\degr$ and the PSG correction for opacity is most accurate at ``small'' solar phase angles.

We used a fixed resolution element, $\Delta\lambda$, that was determined based on the central wavelength of each spectral extract. For 1.80 $\mu$m $\leq \lambda \leq$ 3.2 $\mu$m, we set $\Delta\lambda$ = 0.79 nm, and for 3.2 $\mu$m $< \lambda \leq $ 5.10 $\mu$m, we set $\Delta\lambda$ = 1.32 nm when analyzing NIRSpec data. These values are in good agreement with those from curves for dispersion as a function of wavelength provided by the Space Telescope Science Institute \footnote{\url{https://jwst-docs.stsci.edu/jwst-near-infrared-spectrograph/nirspec-instrumentation/nirspec-dispersers-and-filters\#gsc.tab=0}}. For MIRI, we calculated $\Delta\lambda$ as a function of central extract wavelength following the relationship in \cite{Pontoppidan2024}. We assumed the gas expansion speed varied with \rh{} as $v_\mathrm{exp}=0.8r_H^{-0.5}$ \kms{} \citep{Biver1999,Ootsubo2012}, which gives a value of 0.69 \kms{}. This is in good agreement with the average $v_{\mathrm{exp}}=0.68$ \kms{} measured by \cite{Biver2024a} for E3 using velocity-resolved radio spectra obtained with the IRAM 30 m telescope on 2023 February $2-6$. Uncertainties on the derived parameters were retrieved from the diagonal elements of the covariance matrix, scaled by the square root of the reduced $\chi^2$ statistic of the fit.

Several of the strongest \ce{H2O} emissions in the 2.9 \um{} hot bands measured with the G395H grating are strongly blended with OH$^*$. In particular, emissions from OH$^*$ transitions near 2.886, 2.90, 2.923, and 2.934 \um{} are under-fit by the PSG model. This elevates the derived $N$ and \trot{} for \ce{H2O} as the model attempts to account for the unfit OH$^*$ emission by over-inflating the \ce{H2O}. In contrast, OH$^*$ lines near 2.96 and 2.97 \um{} are better fit. We masked the former OH$^*$ lines when analyzing the 2.9 \um{} hot bands, which produced values for both $N$ and \trot{} that are in better agreement with those derived from the remaining \ce{H2O} bands. Similarly, we masked the positions of the CO and OCS bands when analyzing the 4.5 \um{} \ce{H2O} hot bands.

In addition to providing the overall \ce{H2O} column density, the \ce{H2O} $\nu_1+\nu_3$ (2.6 \um{}) and $\nu_6$ (6 \um{}) bands include multiple strong, unblended lines of its nuclear spin isomers, ortho- and para-\ce{H2O}, enabling us to generate maps of $N(o-\ce{H2O})$ and $N(p-\ce{H2O})$. We used these to create a map of the ortho-to-para ratio (OPR) for \ce{H2O}. The NASA PSG generates $g$-factors for ortho- and para-\ce{H2O} assuming the statistical equilibrium value of 3 regardless of \trot{} \citep{Villanueva2025}. Thus, $\mathrm{OPR}= 3\times N(o-\ce{H2O}) / N(p-\ce{H2O})$ as modeled with the PSG. Similarly, we used the maps of \ce{^12CO2} and \ce{^13CO2} to produce maps of the \ce{^12C}/\ce{^13C} ratio in the coma, where the PSG calculates the intensity of \ce{^13CO2} lines after correcting for its telluric abundance. Thus, $\ce{^12C}/\ce{^13C}=N(\ce{^12CO2}) / N(\ce{^13CO2}) / 89$ as returned by the PSG.

\section{Results} \label{sec:results}
\begin{figure}
\plotone{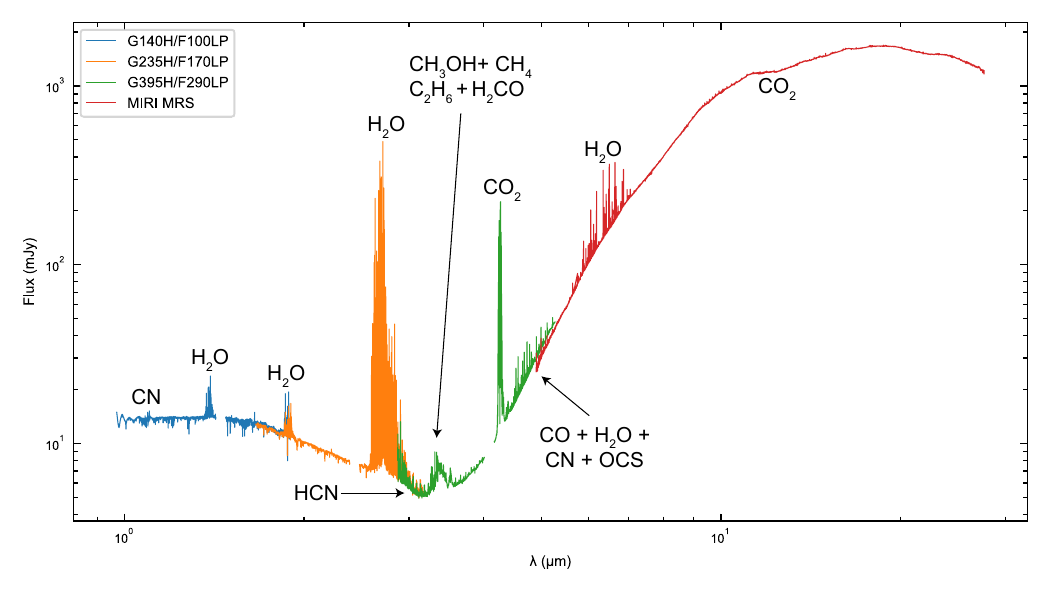}
\caption{JWST spectrum of C/2022 E3 extracted in a 1\farcs3 diameter aperture centered on the nucleus position. Major species are labeled. 
\label{fig:all}}
\end{figure}

\begin{figure*}
\gridline{\fig{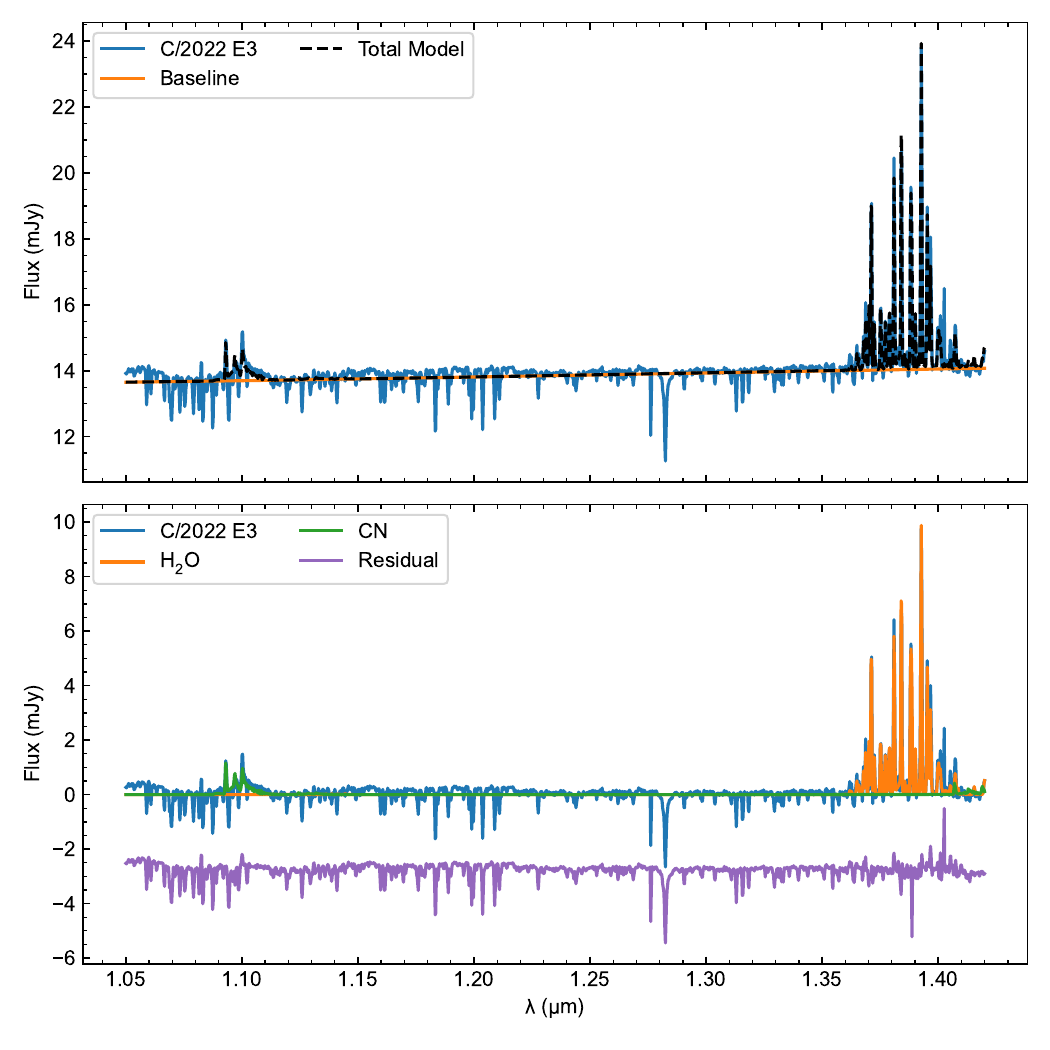}{0.45\textwidth}{(A)}
          \fig{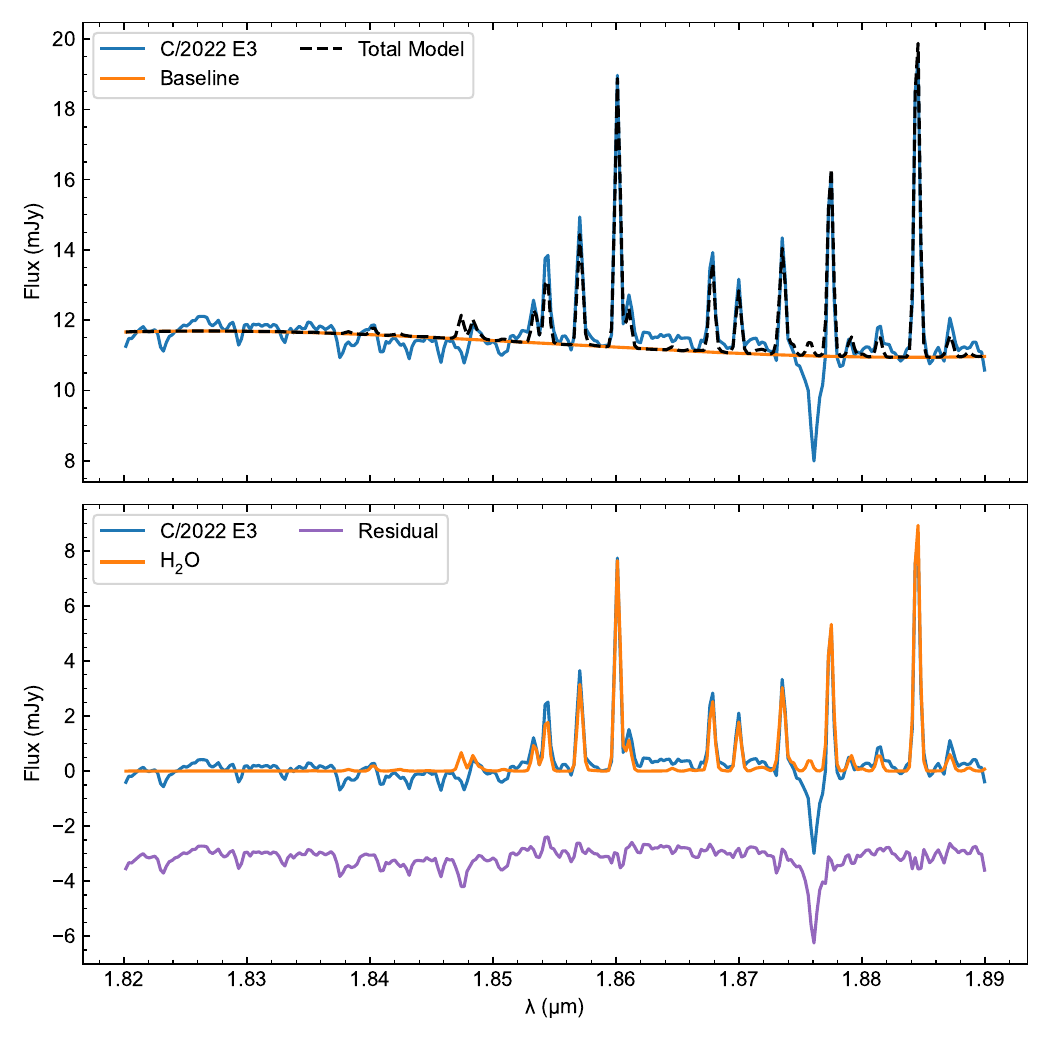}{0.45\textwidth}{(B)}
}
\gridline{\fig{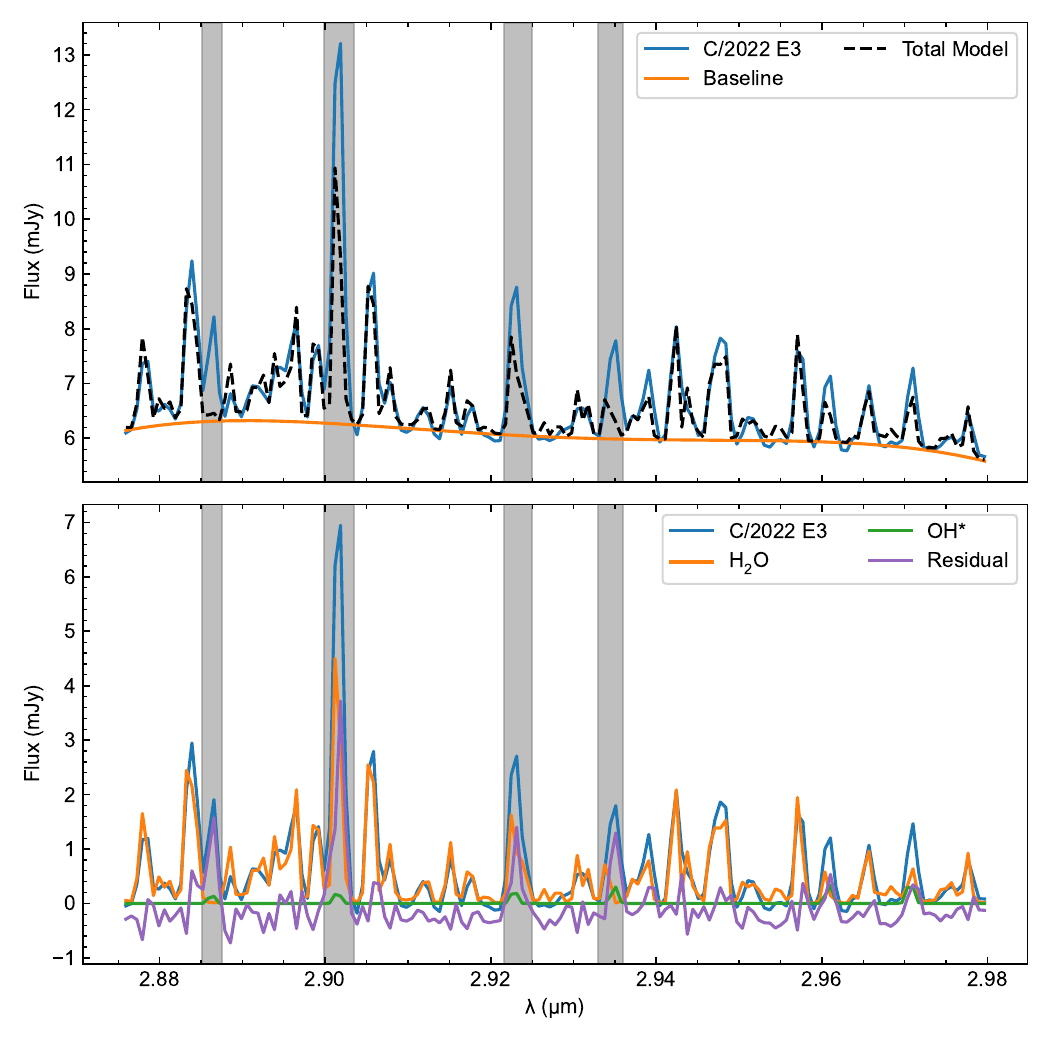}{0.45\textwidth}{(C)}
}
\caption{\textbf{(A)--(C).}  Spectra of comet E3 extracted in a 1\farcs3 diameter aperture and showing detections of \ce{H2O}, OH$^*$, and CN on March 1 with the NIRSpec G140H and G395H gratings. For each figure, the upper panel shows the observed spectrum, spectral baseline, and total fluorescence model. The lower panel shows the baseline-subtracted spectrum and individual best-fit molecular fluorescence models. The residual spectrum (observed$-$baseline$-$models) is shown for comparison and offset vertically. The gray shaded regions in Panel C show regions of OH$^*$ emission that were not reproduced well by the PSG and were masked during fitting.
\label{fig:h2o-panels}}
\end{figure*}

\begin{figure*}
\gridline{\fig{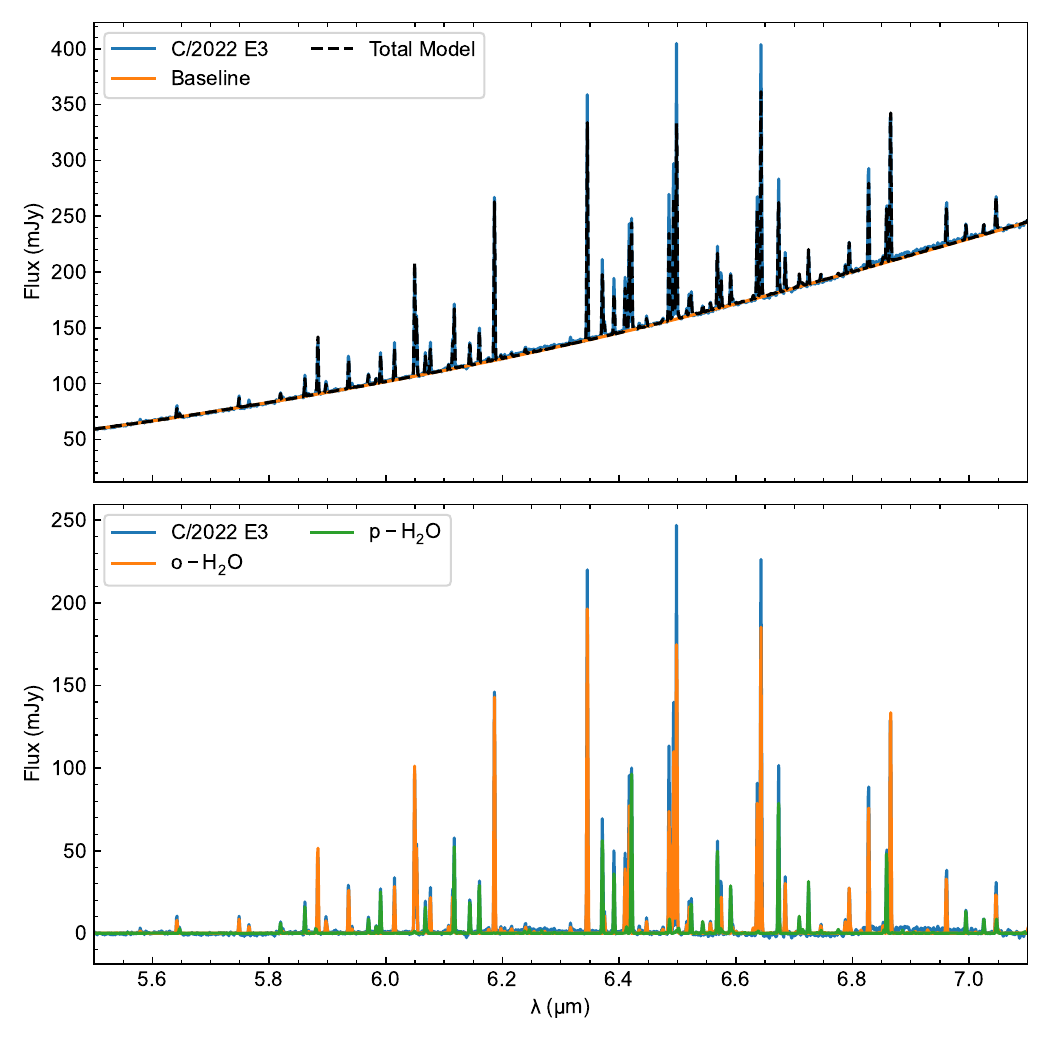}{0.45\textwidth}{(A)}
          \fig{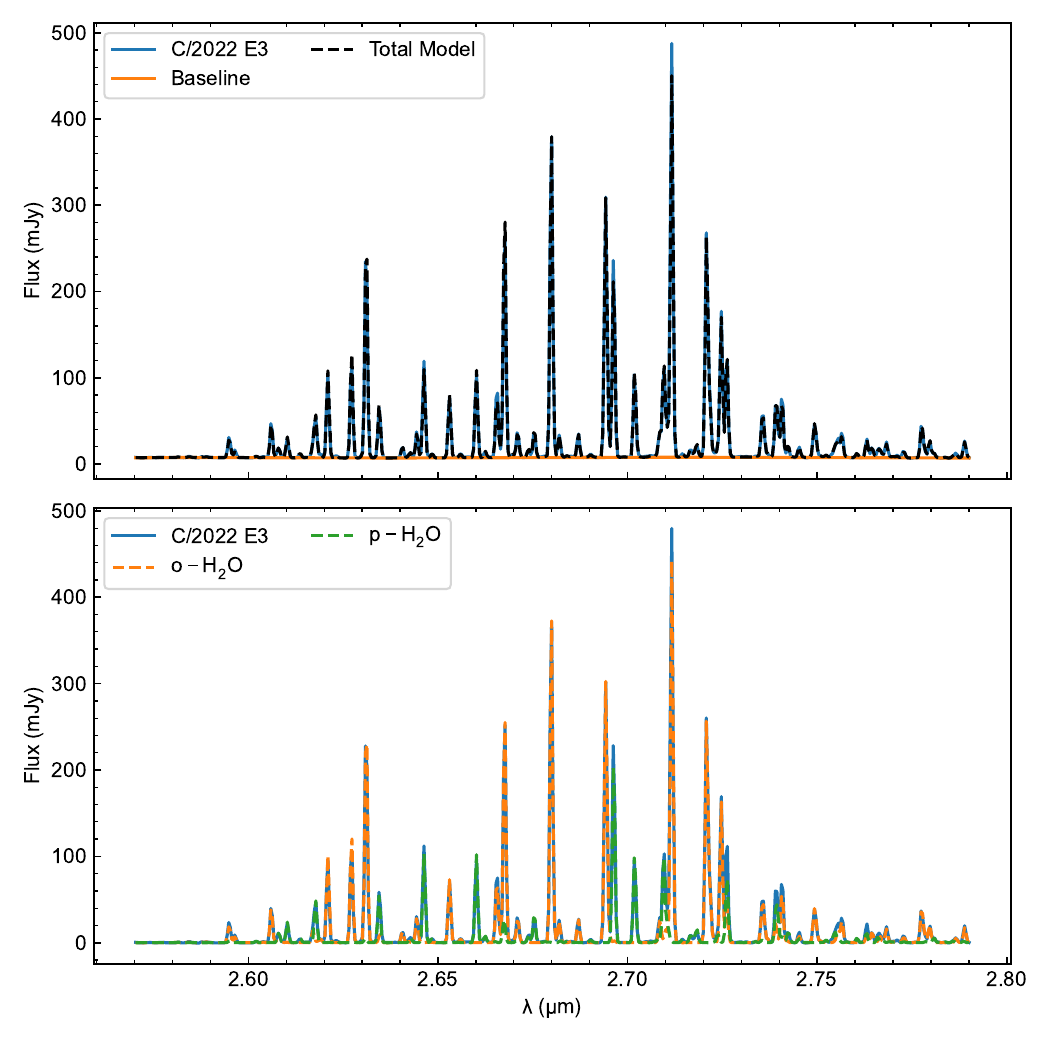}{0.45\textwidth}{(B)}
}
\caption{\textbf{(A)--(B).} As in Figure~\ref{fig:h2o-panels} but for MIRI on February 28 and the NIRSpec G235H grating on March 1. The contributions of $o-\ce{H2O}$ and $p-\ce{H2O}$ are shown separately.
\label{fig:h2o-opr-panels}}
\end{figure*}

\begin{figure*}
\gridline{\fig{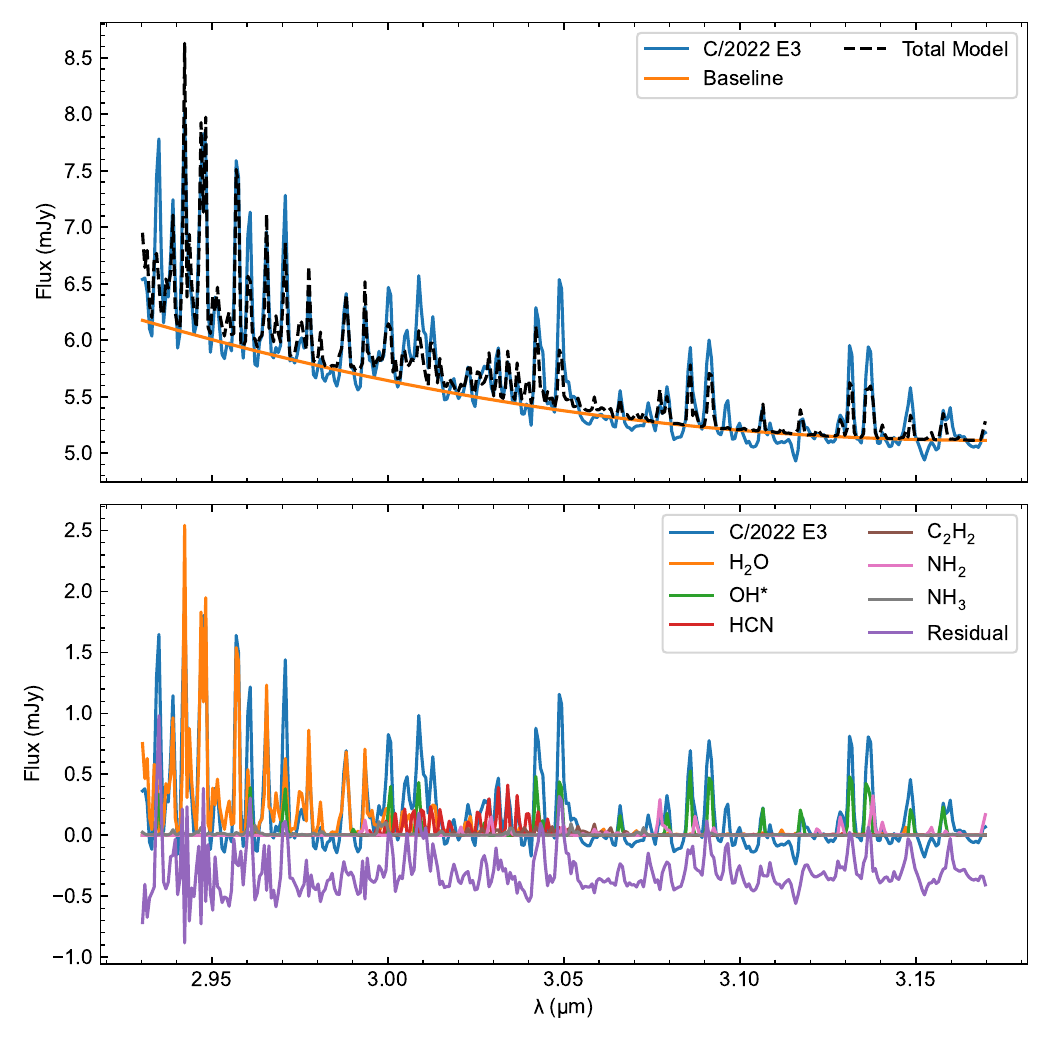}{0.45\textwidth}{(A)}
          \fig{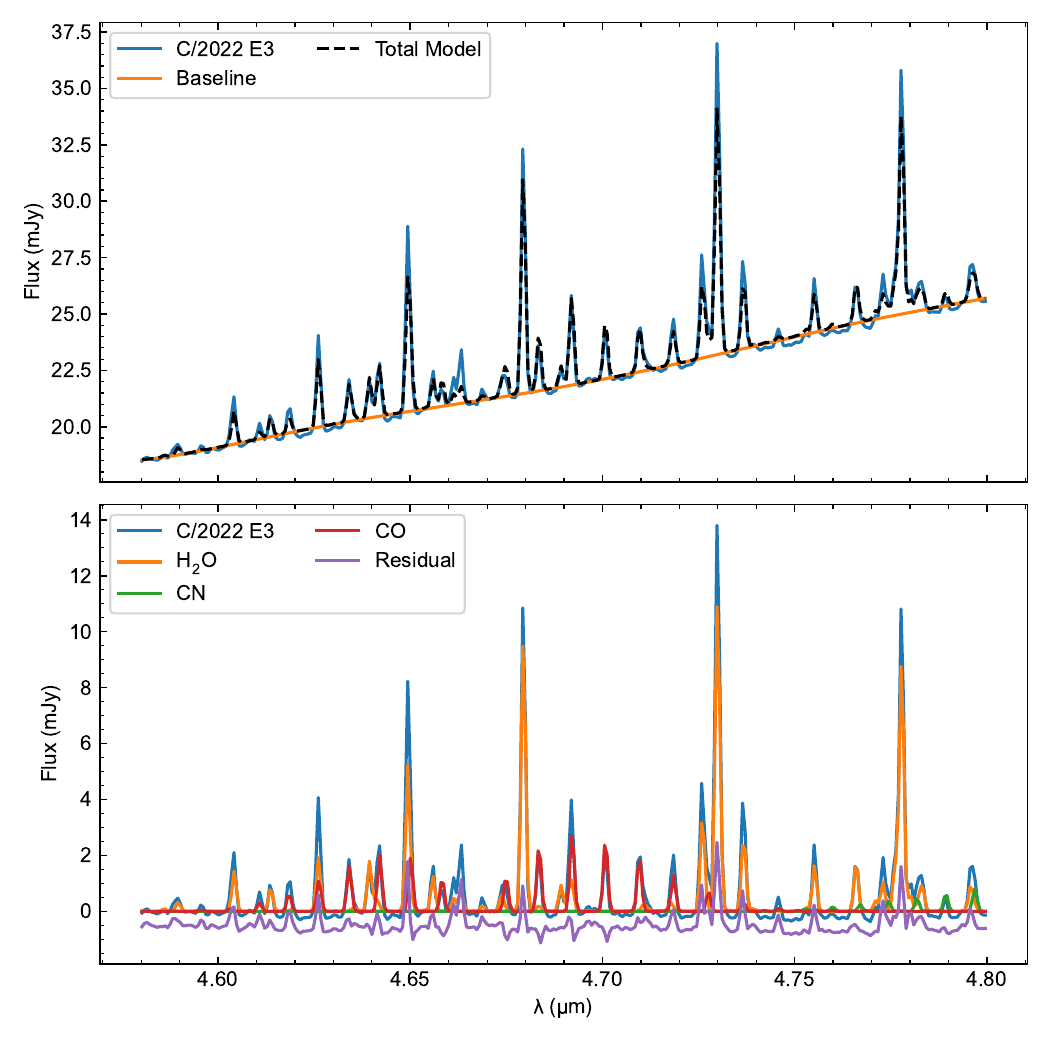}{0.45\textwidth}{(B)}
}
\gridline{\fig{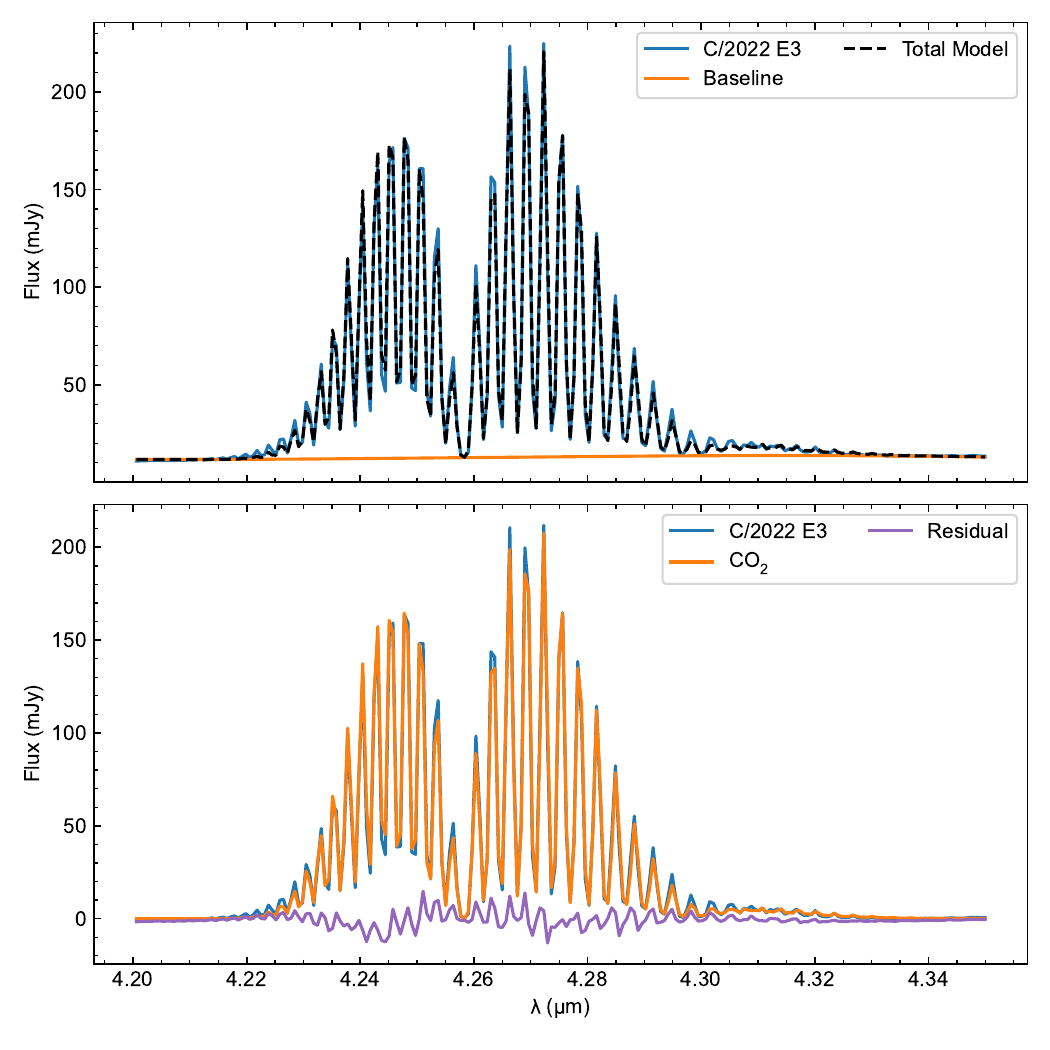}{0.45\textwidth}{(C)}
          \fig{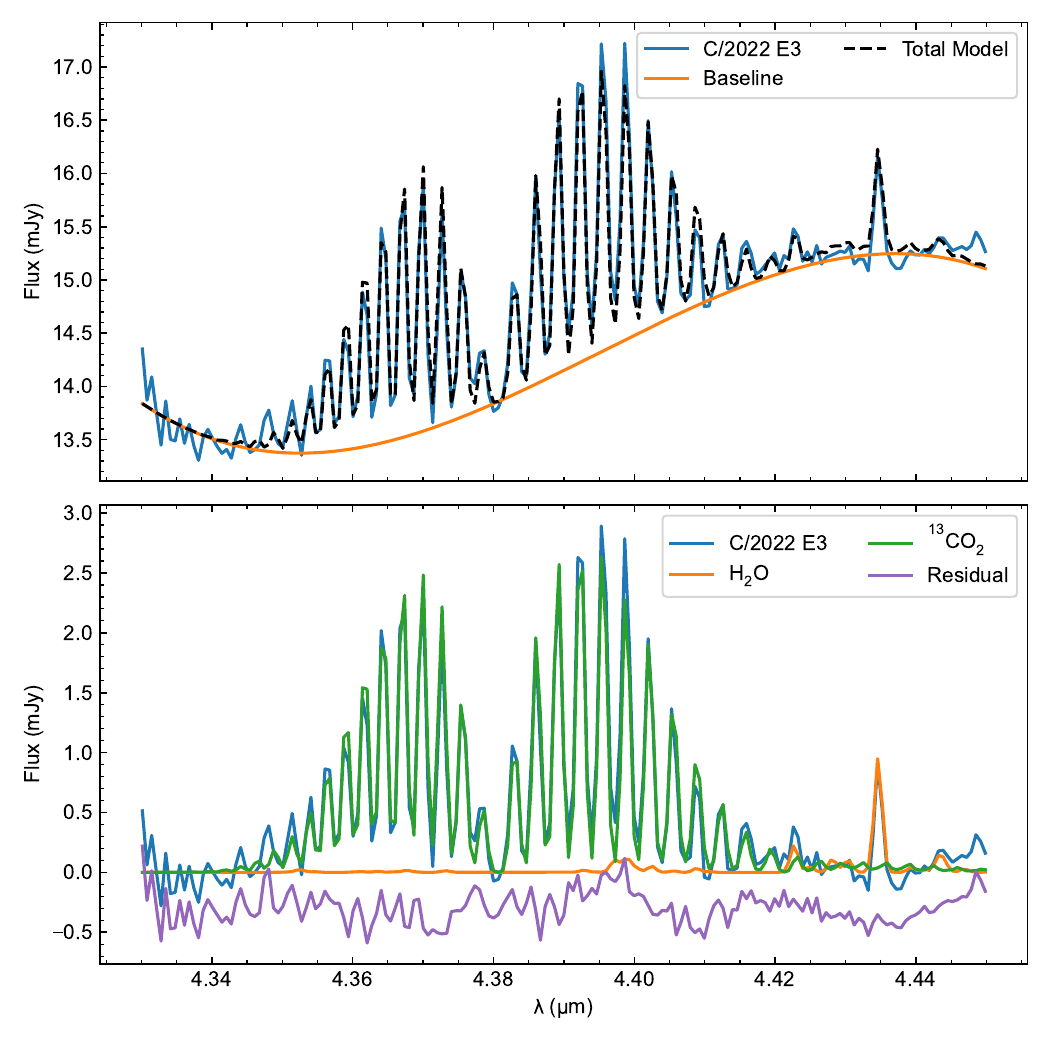}{0.45\textwidth}{(D)}
}
\caption{\textbf{(A)--(D).}  As in Figure~\ref{fig:h2o-panels} for \ce{H2O}, HCN, CO, CN, \ce{CO2}, and \ce{^13CO2} measured with the G395H grating on March 1.
\label{fig:spec-panels1}}
\end{figure*}

\begin{figure*}
\gridline{\fig{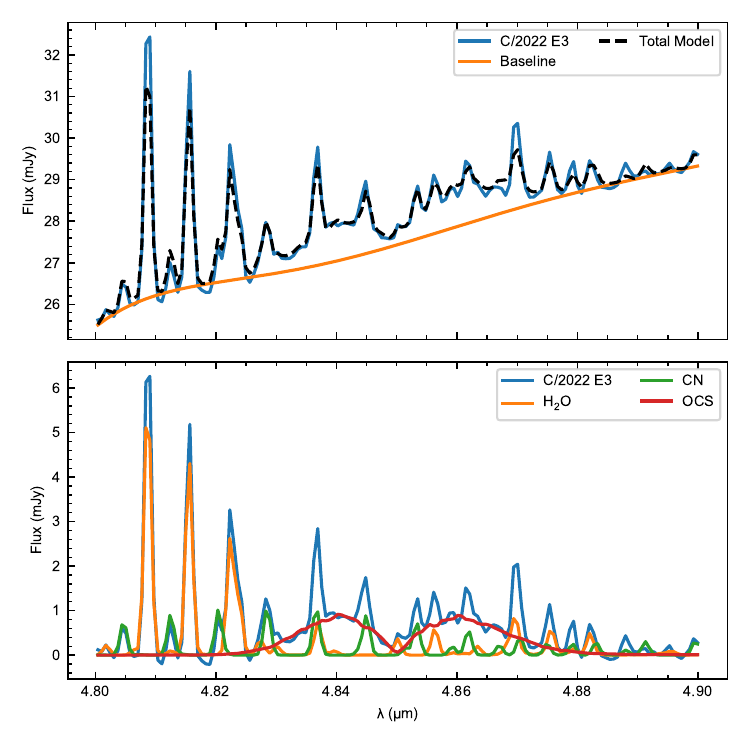}{0.45\textwidth}{(A)}
          \fig{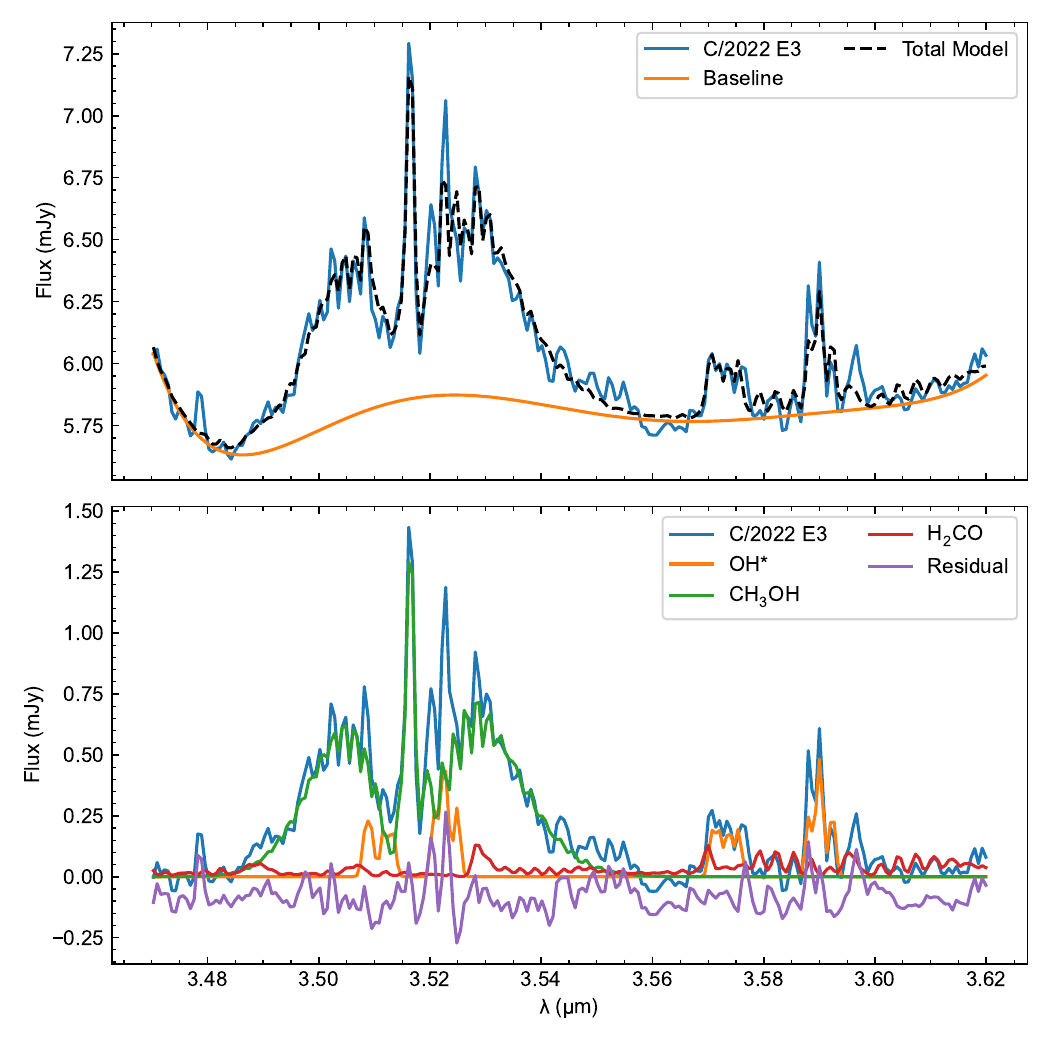}{0.45\textwidth}{(B)}
}
\gridline{\fig{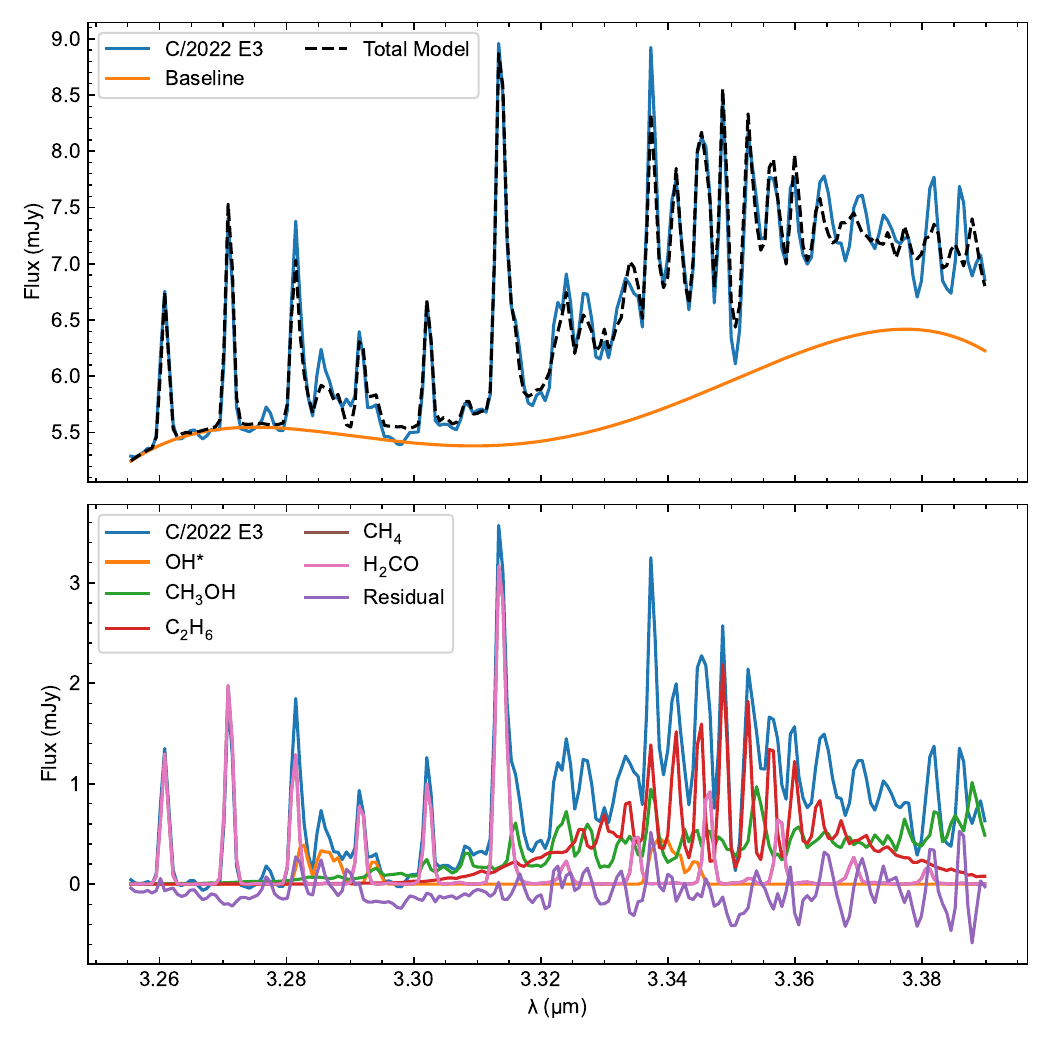}{0.45\textwidth}{(C)}
          \fig{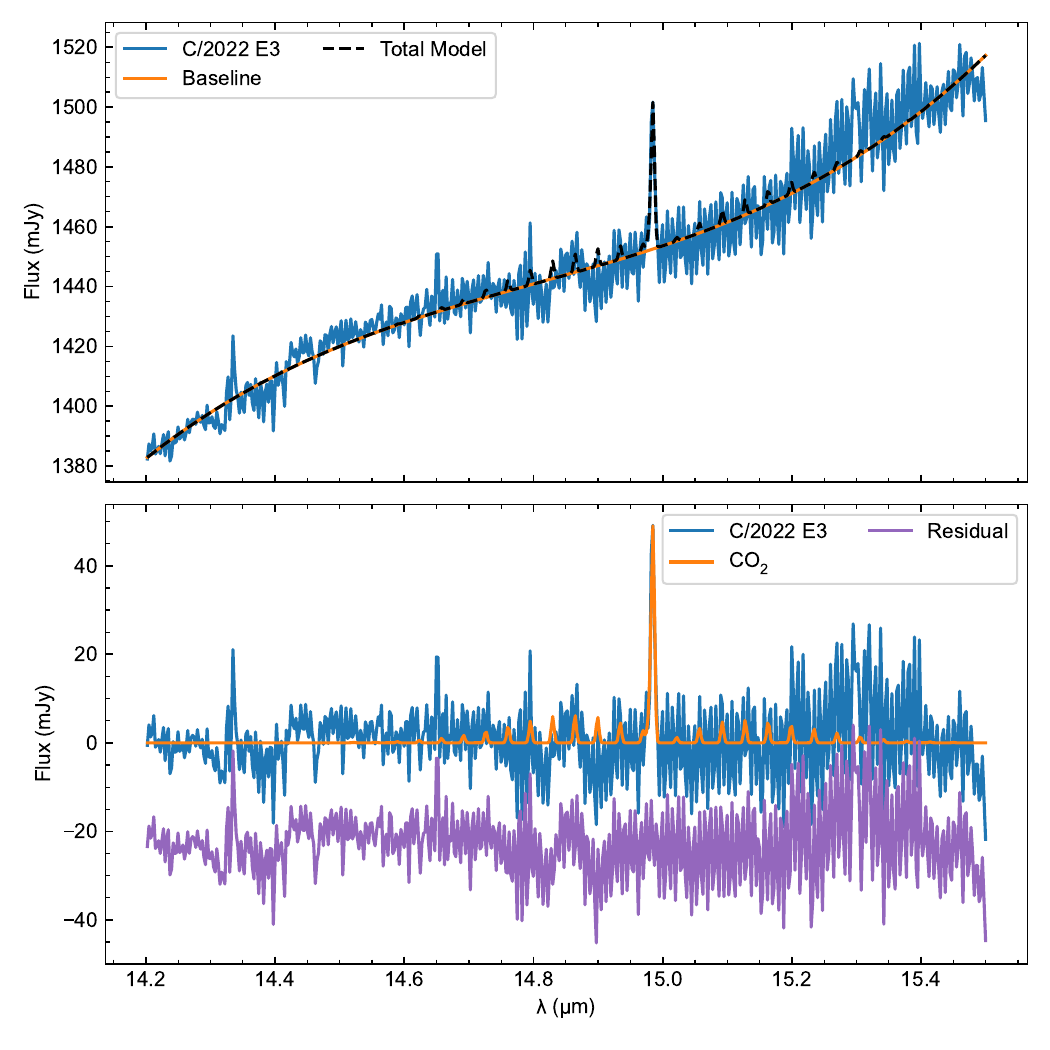}{0.45\textwidth}{(D)}
}
\caption{\textbf{(A)--(C).} As in Figure~\ref{fig:h2o-panels} for \ce{H2O}, OCS, CN, \ce{CH3OH}, \ce{H2CO}, \ce{C2H6}, \ce{CH4}, and OH* measured with the G395H grating on March 1. \textbf{D.} As in Figure~\ref{fig:h2o-panels} for \ce{CO2} measured with MIRI on February 28.
\label{fig:spec-panels2}}
\end{figure*}

A representative JWST spectrum of C/2022 E3 extracted from a nucleus-centered $1\farcs3$ diameter aperture is shown in Figure~\ref{fig:all}. Nucleus-centered molecular production rates ($Q$, \ps{}) were calculated for each species using this spectrum. A corresponding \trot{} in the same aperture was simultaneously retrieved when possible. In general, \trot{} can be reliably constrained when multiple lines sampling a wide range of excitation energies are available for a given species \citep[e.g.,][]{Gibb2012}. These conditions were satisfied for all molecules except \ce{C2H6}, \ce{H2CO}, CN, \ce{NH2}, \ce{NH3}, and \ce{C2H2}. For these species we assumed \trot{} measured for other molecules within the same spectral extract. We assummed \trot{}(\ce{C2H6}) = \trot{}(\ce{CH4}) and posit that the symmetric hydrocarbons likely shared a common \trot{} (see \S~\ref{subsec:trace-results}). We assumed \trot{}(CN) = \trot{}(OCS) (G395H) or \trot{}(\ce{H2O}) (G140H) depending on which molecule was simultaneously measured in the same spectral extract for each NIRSpec grating. When calculating $3\sigma$ upper limits for $Q(\ce{NH2})$, $Q(\ce{NH3})$, and $Q(\ce{C2H2})$, we set their temperature equal to simultaneously extracted \trot{}(HCN). We used active Sun photodissociation rates from \cite{Huebner2015}. The resulting production rates and a comparison against mean values among measured comets are given in Table~\ref{tab:qs}, and best-fit models for this aperture are shown in Figures~\ref{fig:h2o-panels}, \ref{fig:h2o-opr-panels}, \ref{fig:spec-panels1}, and \ref{fig:spec-panels2}.

\begin{figure}
\plotone{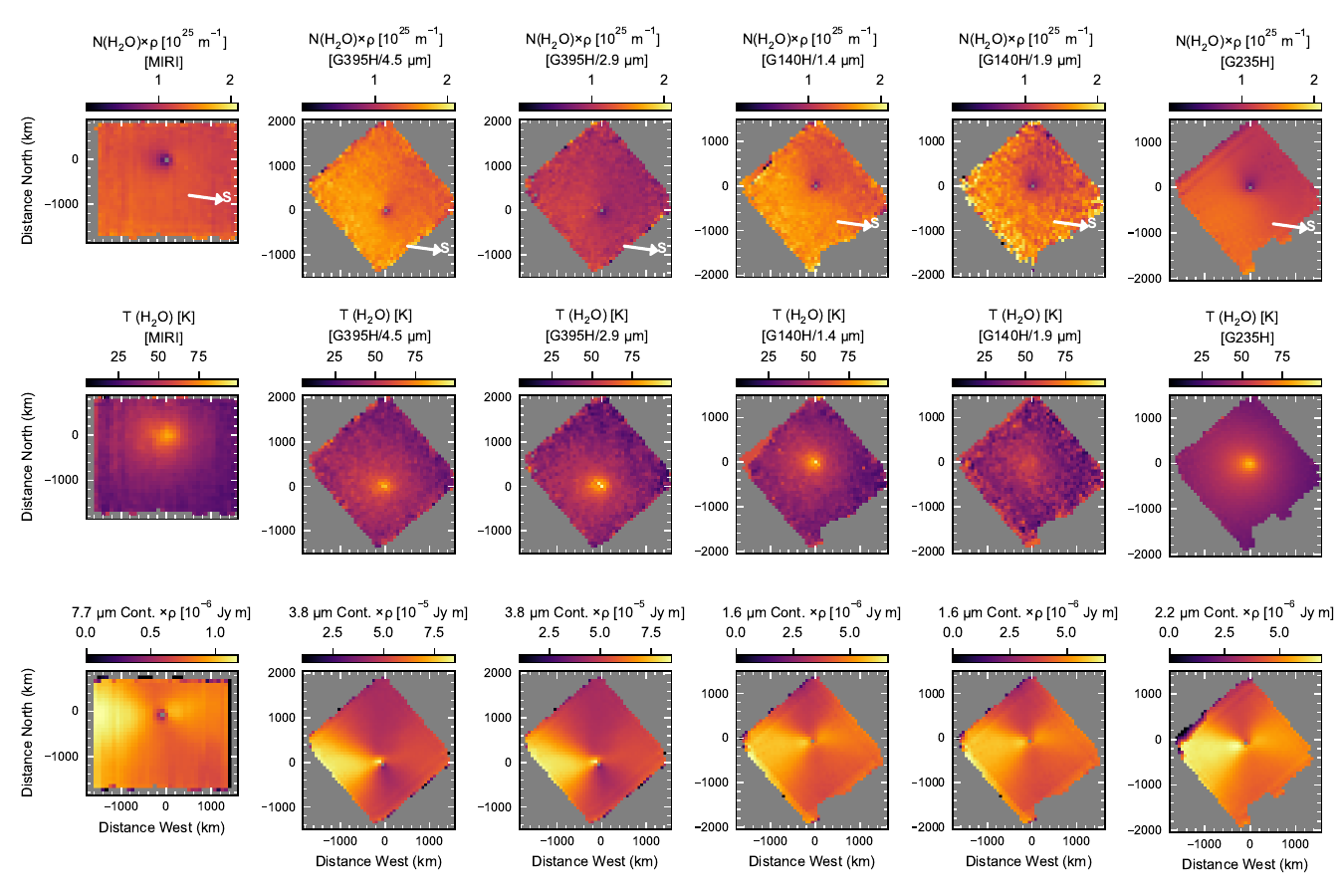}
\caption{\textbf{Upper Panels.} Maps of $N\times\rho$ for \ce{H2O} in chronological order of observation from left to right. The white arrow shows the projected direction of the Sun. Note that the MIRI map was measured on February 28, whereas the remaining maps were constructed from the March 1 observations. \textbf{Middle Panels.} Maps of molecular rotational temperatures. \textbf{Lower Panels.} Maps of continuum intensity $I\times\rho$ simultaneously measured with each \ce{H2O} map. 
\label{fig:water-maps}}
\end{figure}

\begin{figure}
\plotone{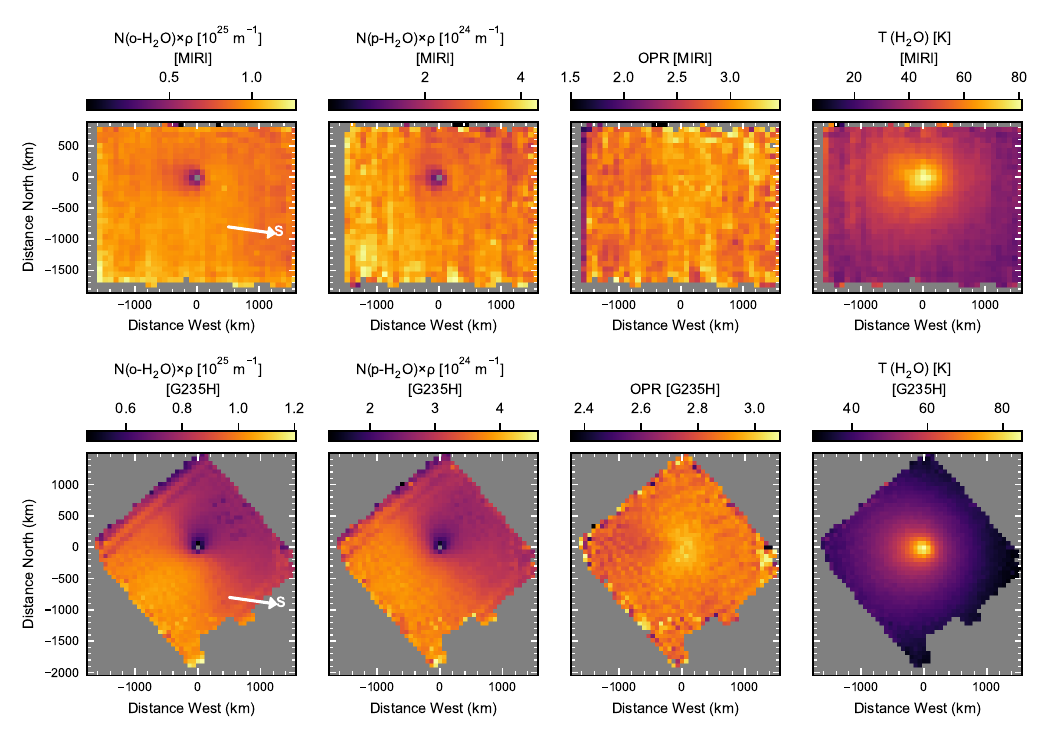}
\caption{\textbf{Upper Panels.} Maps of $N\times\rho$ for ortho-\ce{H2O} and para-\ce{H2O}, the derived OPR, and \trot{}(\ce{H2O}) for the 6 \um{} band measured on February 28. Note that $N$ for each species has been corrected for the derived OPR. \textbf{Lower Panels.} As in the upper panels for the 2.6 \um{} \ce{H2O} bands measured on March 1.
\label{fig:opr-maps}}
\end{figure}

The spaxel-by-spaxel maps for the 1.4, 1.9, 2.6, 2.9, 4.5, and 6 \um{} \ce{H2O} bands and co-measured continuum are shown in Figure~\ref{fig:water-maps}. Maps of $o-\ce{H2O}$, $p-\ce{H2O}$, and OPR derived from the 6 \um{} bands in MIRI and the 2.6 \um{} bands in NIRSpec/G235H are shown in Figure~\ref{fig:opr-maps}, along with histograms of the spaxel-by-spaxel values in Figure~\ref{fig:opr-hist}. Maps for \ce{CO2}, \ce{^13CO2}, CO, \ce{CH4}, \ce{CH3OH}, and OCS are shown in Figure~\ref{fig:trace-maps1}. A map of the \ce{^12C}/\ce{^13C} ratio for \ce{CO2} and a histogram of the spaxel-by-spaxel values are shown in Figure~\ref{fig:carbon-ratio}.

\begin{deluxetable*}{ccccc}
\tablenum{2}
\tablecaption{\label{tab:qs} Nucleus-Centered Molecular Production Rates and Rotational Temperatures in C/2022 E3 and Comets Measured}
\tablewidth{0pt}
\tablehead{
\colhead{Molecule} & \colhead{$Q$} & \colhead{\trot{}} & \colhead{$Q_x/Q(\ce{H2O})$} & \colhead{$\langle Q_x/Q(\ce{H2O})\rangle$}  \\
\colhead{} & \colhead{($10^{26}$ \ps{})} & \colhead{(K)} & \colhead{(\%)} & \colhead{(\%)} 
}
\startdata
\multicolumn{5}{c}{2023 February 28, MIRI} \\
\ce{H2O} & $256\pm16$ & $63.5\pm0.7$ & 100 & 100 \\
\ce{CO2} & $46.0\pm6.1$ & (63) & $18.0\pm2.4$ & $4-30$ \\
\hline
\multicolumn{5}{c}{2023 March 1, G395H/F290LP} \\
\ce{H2O} & $248\pm9$ &  $63\pm2$ & 100 & 100  \\
\ce{CO2} & $28.1\pm1.5$ & $71.3\pm0.7$ & $10.5\pm0.1$ & $4-30$  \\
\ce{^13CO2} & $0.35\pm0.02$ & $73\pm2$ & $0.131\pm0.003$ & ...   \\
CO & $1.98\pm0.15$ & $39\pm2$ & $0.74\pm0.04$ & $0.3-26$ \\
OCS & $0.17\pm0.03$ & $57\pm4$ & $0.062\pm0.009$ & $0.04-0.4$ \\
\ce{CH4} & $1.40\pm0.08$ & $72\pm3$ & $0.52\pm0.02$ & $0.15-2.7$ \\
\ce{C2H6} & $1.55\pm0.12$ & (72) & $0.58\pm0.03$ & $0.1-2.7$ \\
\ce{CH3OH} & $3.36\pm0.22$ & $59\pm2$ & $1.25\pm0.05$ & $<0.13-4.3$ \\
\ce{H2CO} & $0.17\pm0.03$ & (59) & $0.062\pm0.013$ & $<0.02-1.1$ \\
HCN & $2.96\pm0.06$ & $62\pm4$ & $0.11\pm0.02$ & $0.03-0.5$ \\
CN & $0.13\pm0.01$ & (58) & $0.050\pm0.005$ & $<0.1-1$ \\
\ce{NH2} & $<1.2$ $(3\sigma)$ & (62) & $<0.4$ $(3\sigma)$ & $<0.1-1$ \\
\ce{NH3} & $<1.6$ $(3\sigma)$ & (62) & $<0.6$ $(3\sigma)$ & $0.1-3.6$ \\
\ce{C2H2} & $<0.2$ $(3\sigma)$ & (62) & $<0.08$ $(3\sigma)$ & $0.03-0.37$ \\
\hline
\multicolumn{5}{c}{2023 March 1, G140H/F100LP} \\
\ce{H2O} & $296\pm10$ & $51\pm3$ & 100 & 100 \\
CN & $0.12\pm0.01$ & (51) & $0.040\pm0.004$ & $<0.1-1$ \\
\hline
\multicolumn{5}{c}{2023 March 1, G235H/F170LP} \\
\ce{H2O} & $270\pm1$ & $62.4\pm0.4$ & 100 & 100 \\
\enddata
\tablecomments{$Q$ and \trot{} as derived from fits to the nucleus-centered 1\farcs3-diameter spectrum. Values in parentheses are assumed. $\langle Q_x/Q(\ce{H2O})\rangle$ are the ranges of molecular abundances in comets measured \citep{Biver2024b}. Values for CN and \ce{NH2} are given as ranges across the comet population, which are often reported as mixing ratios relative to OH rather than \ce{H2O} \citep{Biver2024b}.}
\end{deluxetable*}

\subsection{Analysis of \ce{H2O} Emission}\label{subsec:h2o-results}

Expansion dilution of the coma leaves trends in gas distribution difficult to discern in plots of $N$. Plotting $N\times\rho$, where $\rho$ (m) is the radial distance from the nucleus, corrects for this effect and more readily reveals coma anisotropies: isotropic outgassing in the absence of photolysis and coma acceleration would produce a flat $N\times\rho$ dependence with nucleocentric distance. Significant anisotropies are evident in the spatial distributions of column density and \trot{} among the species in E3's coma. A dual dust jet oriented along the Sun-comet axis is evident in each instrumental setting, although the anti-sunward component is brighter. The \ce{H2O} emission appears relatively symmetric in the MIRI observations, and the apparently higher $N\times\rho$ in the anti-sunward direction during the NIRSpec observations may be manifestations of differences in coma expansion speed ($N\propto Q/v_{\mathrm{exp}}$) in the sunward versus anti-sunward hemispheres of the coma. Indeed, \cite{Biver2024a} found evidence for a higher sunward expansion speed compared to the anti-sunward direction based on IRAM 30 m measurements of E3 approximately one month prior to these JWST observations.

Our results for the 2.6, 2.9, and 4.5 \um{} bands are consistent with the analysis of \cite{Foster2026}, which found a higher \trot{} in the anti-sunward direction. This was attributed to more efficient adiabatic cooling in the sunward direction. This \trot{} asymmetry is also evident for our analysis of the 6 \um{} bands in MIRI, although it is not as clearly evident for the 1.9 \um{} bands in the NIRSpec/G140H grating on visual inspection. However, radial \trot{} profiles demonstrate that the asymmetry is indeed present for all \ce{H2O} bands and trace species (\S~\ref{subsec:trace-results}).

\begin{figure}
\plotone{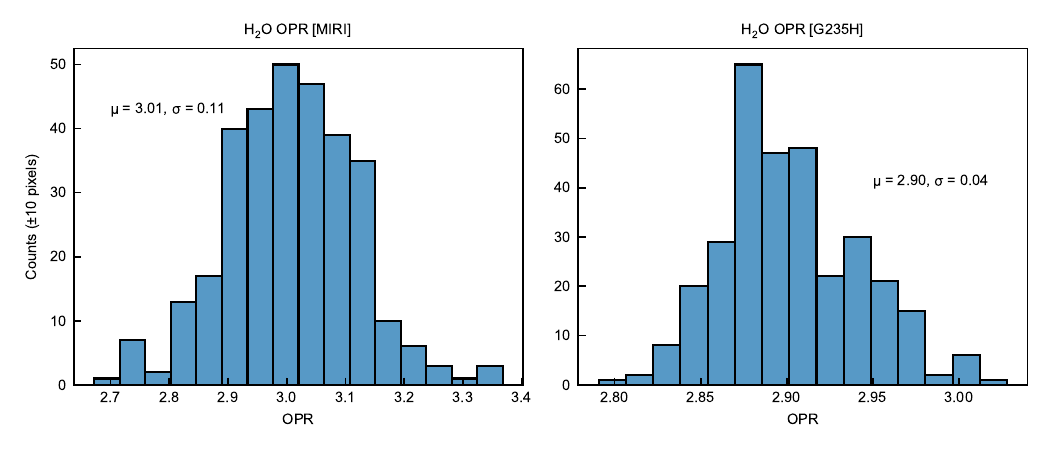}
\caption{Histogram of spaxel-by-spaxel OPR values for \ce{H2O} measured on February 28 and March 1 drawn within a 10-spaxel radius of the comet photocenter.
\label{fig:opr-hist}}
\end{figure}

Our map of the OPR for \ce{H2O} demonstrates an essentially flat value with nucleocentric distance, independent of changes in the total column density or \trot{} for both the 2.6 \um{} bands measured with NIRSpec and the 6 \um{} bands measured with MIRI. We derived a coma-averaged value of $\mathrm{OPR}=3.01\pm0.11$ for the MIRI observations and $\mathrm{OPR}=2.90\pm0.04$ for the NIRSpec observations by calculating statistics for values retrieved within a 10-spaxel radius of the comet photocenter (Figure~\ref{fig:opr-hist}). These values do not show evidence for significant deviation from the statistical value of 3. This is in contrast with other comets measured with JWST such as C/2017 K2 \citep[$\sim2.75$;][]{Woodward2025}, 3I/ATLAS \citep[$2.7\pm0.2$;][]{Roth2026c}, and 81P/Wild 2 \citep[$2.76\pm0.05$;][]{Roth2026d}, which showed lower values for OPR. However, the constancy of OPR across the coma within the JWST field of view is consistent with results for 3I/ATLAS and other solar system comets \citep[e.g.,][]{Bonev2007,BockeleeMorvan2009,Roth2026c}.

\subsection{Analysis of Trace Species}\label{subsec:trace-results}

\begin{figure}
\plotone{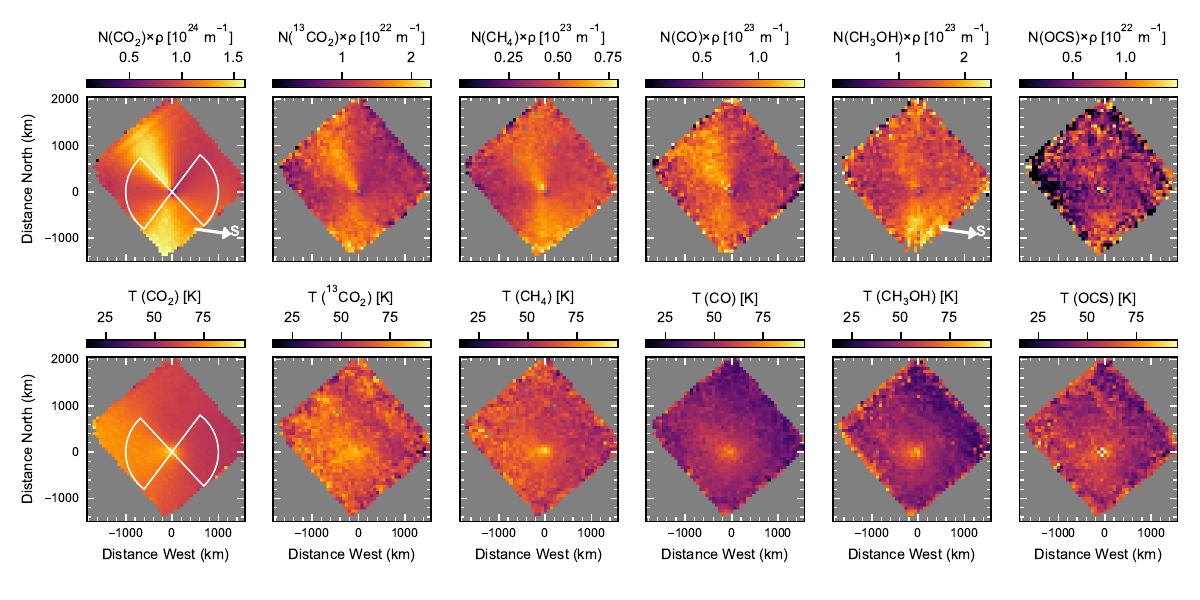}
\caption{\textbf{Upper Panels.} Maps of $N\times\rho$ for \ce{CO2}, \ce{^13CO2},  \ce{CH4}, CO, \ce{CH3OH}, and OCS on March 1. \textbf{Lower Panels.} Maps of molecular rotational temperatures. The white wedge illustrates the anti-sunward- and sunward-facing regions used to analyze the asymmetry in \trot{} for each molecule along the projected Sun-comet vector. Each wedge terminates at $r=1000$ km from the nucleus. 
\label{fig:trace-maps1}}
\end{figure}

Our analysis of trace species emission in E3 shows significant differences in the spatial distributions of $N$ and \trot{} compared to \ce{H2O}. \ce{CO2}, the molecule with the highest SNR, clearly shows a dual jet feature offset to the northeast and southeast from the projected Sun-comet vector (Figure~\ref{fig:trace-maps1}). The same structure is visible, although at lower SNR, for \ce{^13CO2}, \ce{CH4}, CO, and \ce{CH3OH}. The SNR for OCS is too low to draw firm conclusions regarding whether it traces the spatial morphology of \ce{CO2} or \ce{H2O}. Despite the differences in $N\times\rho$, maps of \trot{} for the trace species are consistent with the trend for \ce{H2O} of a warmer anti-sunward hemisphere. However, the overall magnitude of \trot{} for the trace species varies among themselves and compared to \ce{H2O}.

As noted above, \cite{Foster2026} found that \trot{} for both \ce{H2O} and \ce{CH3OH} were warmer in the anti-sunward direction of the coma compared to the sunward direction based on analysis of spaxels lying within $\pm45$\degr \ of the projected Sun-comet vector. We performed a similar analysis here, analyzing \trot{} at each spaxel in the region of the clear anti-sunward fan-like temperature enhancement for \ce{CO2}, as well as its complement in the sunward direction. Thus, the anti-sunward region is defined as a wedge spanning sky-projected position angles of $43\degr-143\degr$ and the sunward region is its complement (position angles $223\degr-313\degr$) as shown in Figure~\ref{fig:trace-maps1}.

\begin{figure*}
\gridline{\fig{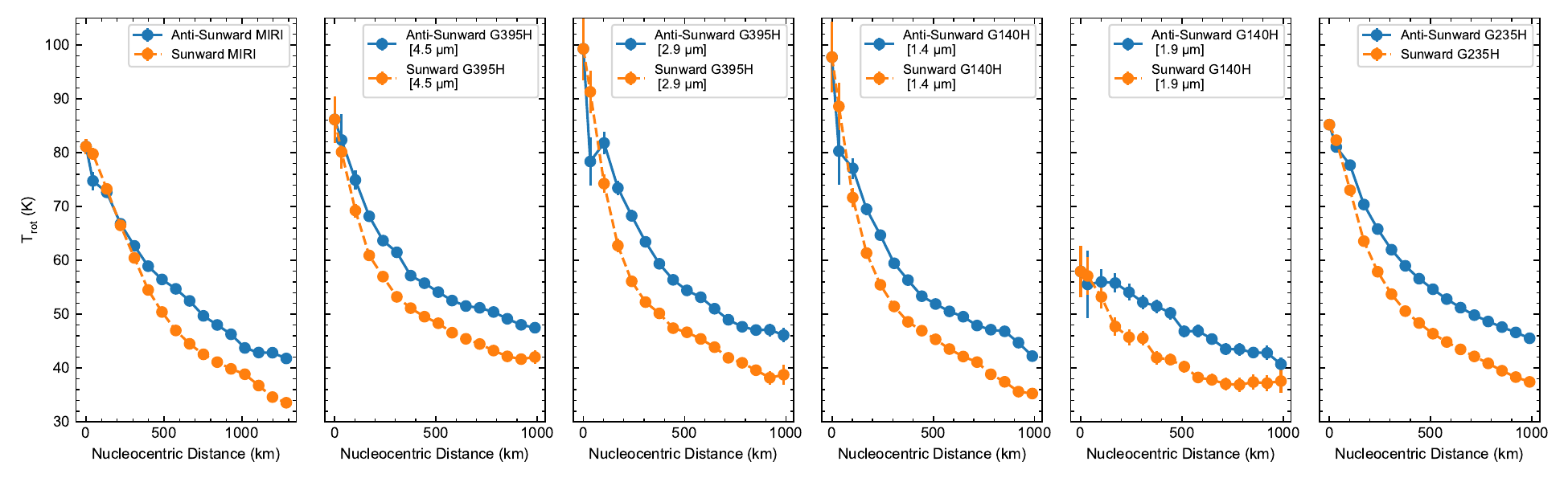}{0.95\textwidth}{(A)}
}
\gridline{\fig{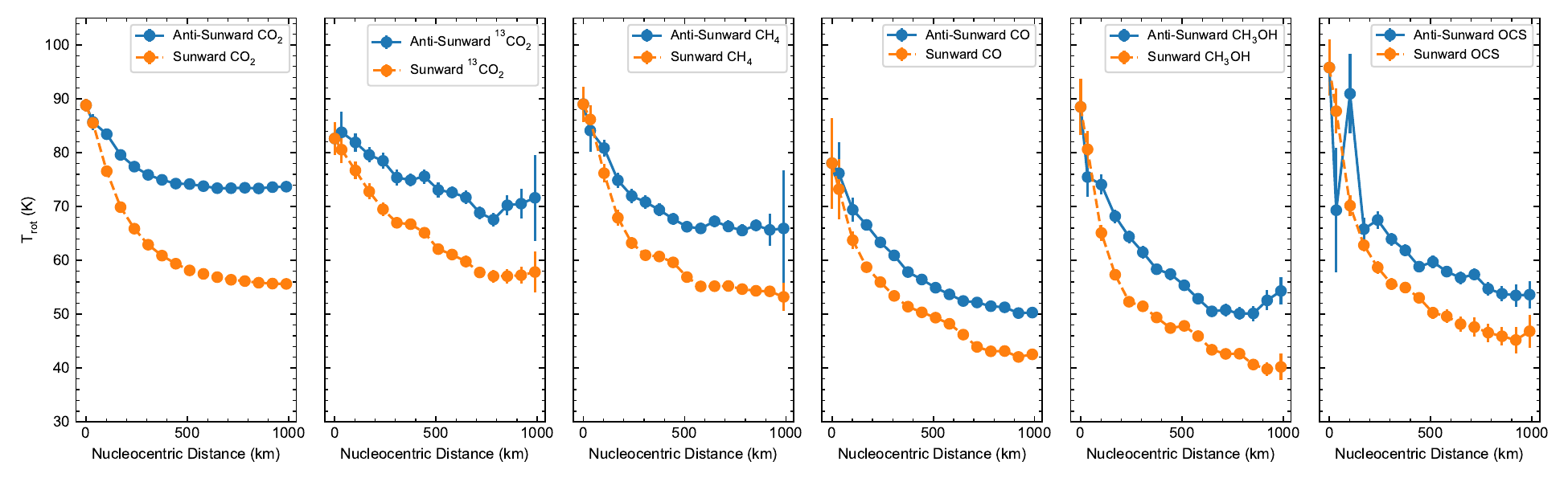}{0.95\textwidth}{(B)}
}
\caption{\textbf{(A).} Azimuthally averaged sunward and anti-sunward radial \trot{} profiles for \ce{H2O} for MIRI and each NIRSpec grating ordered chronologically. \textbf{(B).} Azimuthally averaged radial \trot{} profiles for \ce{CO2}, \ce{^13CO2}, \ce{CH4}, CO, \ce{CH3OH}, and OCS.
\label{fig:tcurves}}
\end{figure*}

We then created azimuthally averaged radial profiles of \trot{} in each region for each molecule out to $r=1000$ km nucleocentric distance. Our results are shown in Figure~\ref{fig:tcurves}. The results are consistent with the analysis by \cite{Foster2026} for \ce{H2O} and \ce{CH3OH}, and demonstrate that the sunward/anti-sunward dichotomy in cooling is apparent for all detected molecules with sufficient off-nucleus SNR to map \trot{}. For \ce{H2O}, our analysis of the 1.9 \um{} bands \citep[not considered by][]{Foster2026} shows the lowest average \trot{} in the dataset, with a nucleus-centered temperature of only $\sim60$ K compared to $\sim80-90$ K for the other bands. However, the \trot{} attained at significant distances from the nucleus is consistent among all bands ($\sim40$ K). These 1.9 \um{} \ce{H2O} features are associated with overtone and combination bands (e.g., $\nu_1+\nu_2$, $\nu_1+2\nu_2$, $\nu_2+2\nu_3$) and have the lowest opacity of any \ce{H2O} band measured in this study. Discounting the central value (where the very inner pixels near the nucleus are heavily affected by the PSF), the higher \trot{} in the inner coma for the 1.4, 2.6, 2.9, 4.5, and 6 \um{} bands compared to the 1.9 \um{} band can be understood by the higher opacities of the former. For optically thick lines, the derived \trot{} is an ``apparent'' \trot{}, which is inherently higher than the true \trot{} that describes the population distribution in the ground state. The strongest lines originate from lower energy levels, which are much more attenuated by optical depth effects compared to the fainter lines that originate from higher ro-vibrational energy levels; thus, the ``apparent'' \trot{} under optically thick conditions is skewed towards higher values \citep{Cheng2022,Debout2016}. The general consistency of all \ce{H2O} bands at larger nucleocentric distances indicates that opacity effects are significantly reduced in the outer coma, and that the 1.9 \um{} band gives the most direct measure of \trot{} for \ce{H2O} in the inner coma. 

On the other hand, the trace species follow a more predictable pattern: species with larger dipole moments radiatively cool more quickly than those lacking a dipole. For instance, \trot{}(\ce{CO2}) is $\sim15$ K warmer in both the anti-sunward and sunward directions than \ce{CH3OH} at 1000 km radius from the nucleus. These trends have been predicted by radiative transfer calculations and demonstrated in previous JWST spatial-spectral studies of comets \citep[e.g.,][]{Bodewits2024,Woodward2025,Roth2026c}.

\begin{figure}
\plotone{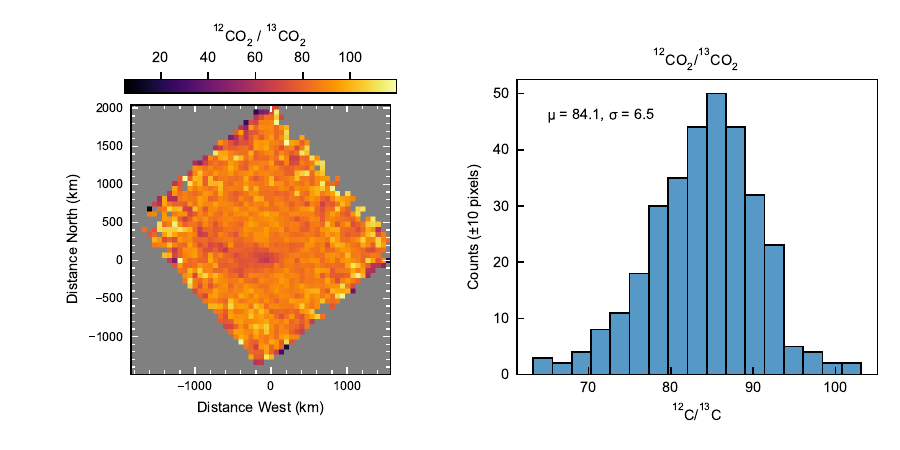}
\caption{\textbf{Left.} Map of \ce{^12C}/\ce{^13C} for \ce{CO2} measured on March 1. \textbf{Right.} Histogram of spaxel-by-spaxel \ce{^12C}/\ce{^13C} values drawn within a 10-spaxel radius of the comet photocenter.
\label{fig:carbon-ratio}}
\end{figure}

Similar to our analysis of spatial variation in the \ce{H2O} OPR with nucleocentric distance, we mapped the variation of \ce{^12C}/\ce{^13C} from our simultaneous measures of \ce{^12CO2} and \ce{^13CO2} and created a histogram of the spaxel-by-spaxel values within a 10-spaxel radius of the nucleus (Figure~\ref{fig:carbon-ratio}). Our results show a flat \ce{^12C}/\ce{^13C} ratio throughout the coma, and the coma-averaged value of $84.1\pm6.5$ is consistent with the terrestrial value of 89. This is consistent with the volatile carbon reservoir measured across the solar system, with interstellar comet 3I/ATLAS being the only object measured to date whose \ce{^12C/^13C} ratio deviated significantly from terrestrial \citep[being 141-191 for \ce{CO2};][]{Cordiner2026}.

In terms of molecular abundances of the trace species, Table~\ref{tab:qs} shows that E3 was consistent with mean values in measured comets for \ce{CO2} and \ce{C2H6} yet depleted for all other molecules \citep{Biver2024b}. E3 is among multiple comets measured at near-infrared wavelengths with such a ``mixed'' composition \citep[consistent with the average in some molecules with enriched or depleted others, e.g.,][]{DelloRusso2014,Roth2017,Faggi2018,Saki2021,Woodward2025}.

\ce{CO2} presents an interesting case, having been measured through its 15 \um{} hot bands with MIRI on February 28 and through its $\nu_3$ fundamental and $\nu_1+\nu_3-\nu_1$ hot bands with NIRSpec on March 1. Table~\ref{tab:qs} shows that \ce{CO2}/\ce{H2O} fell from 18\% (moderately enriched) on February 28 to 10.5\% (consistent with mean cometary values) on March 1. Indeed, $Q(\ce{H2O})$ decreased by only 3\% between these epochs, and the bulk of the change in \ce{CO2}/\ce{H2O} can be attributed to a decrease in $Q(\ce{CO2})$ by 48\% from the MIRI measurements to the NIRSpec observations. With spectral resolution capable of resolving CO from \ce{CO2} in the near-infrared and unparalleled sensitivity in the mid-infrared, JWST is providing routine access to \ce{CO2} in comets at both wavelengths for the first time. Our findings of higher $Q(\ce{CO2})$ measured with MIRI compared to NIRSpec are remarkably similar to results found for 3I/ATLAS \citep{Cordiner2026,Belyakov2026} and 81P/Wild 2 \citep{Roth2026d}. Although it is tempting to find a pattern, all three comets had their MIRI observations taken $\geq20$ hours apart from NIRSpec. Furthermore, 3I/ATLAS in particular was undergoing steep changes in activity level and abundance ratios on timescales of days as it receded from the Sun and crossed the \ce{H2O} ice line near $2-3$ au \citep{Cordiner2026,Belyakov2026,Salazar2026}. Thus, a coincidence of significant temporal variability in $Q(\ce{CO2})$ in all three comets cannot be ruled out. Future investigations with nearly simultaneous MIRI and NIRSpec measurements of cometary \ce{CO2} may help to disentangle outgassing variability from potential modeling effects.

\subsection{Analysis of Mid-Infrared OH$^*$ Prompt Emission}\label{subsubsec:miri-oh}
\begin{figure*}
\gridline{\fig{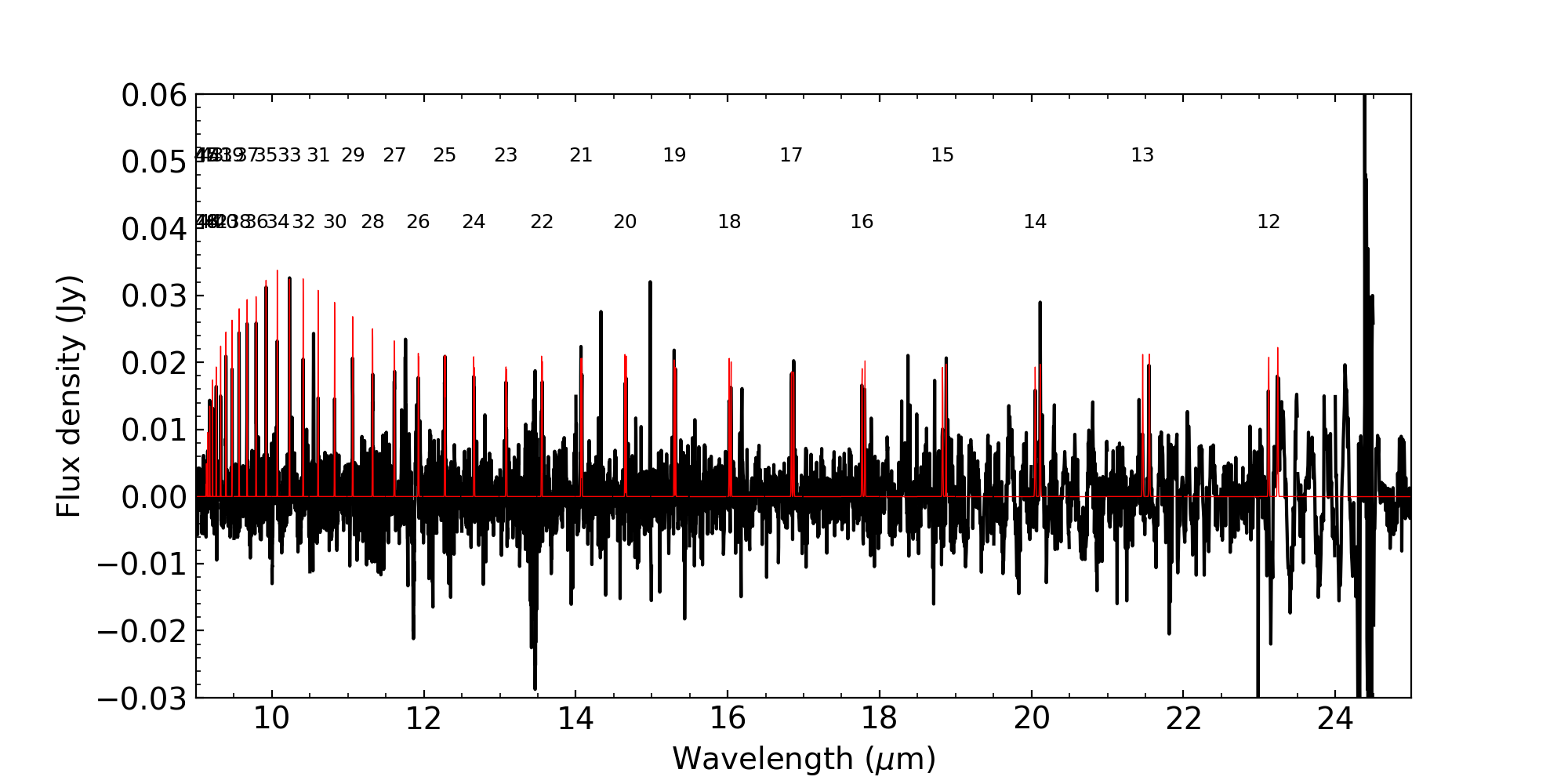}{0.45\textwidth}{(A)}
          \fig{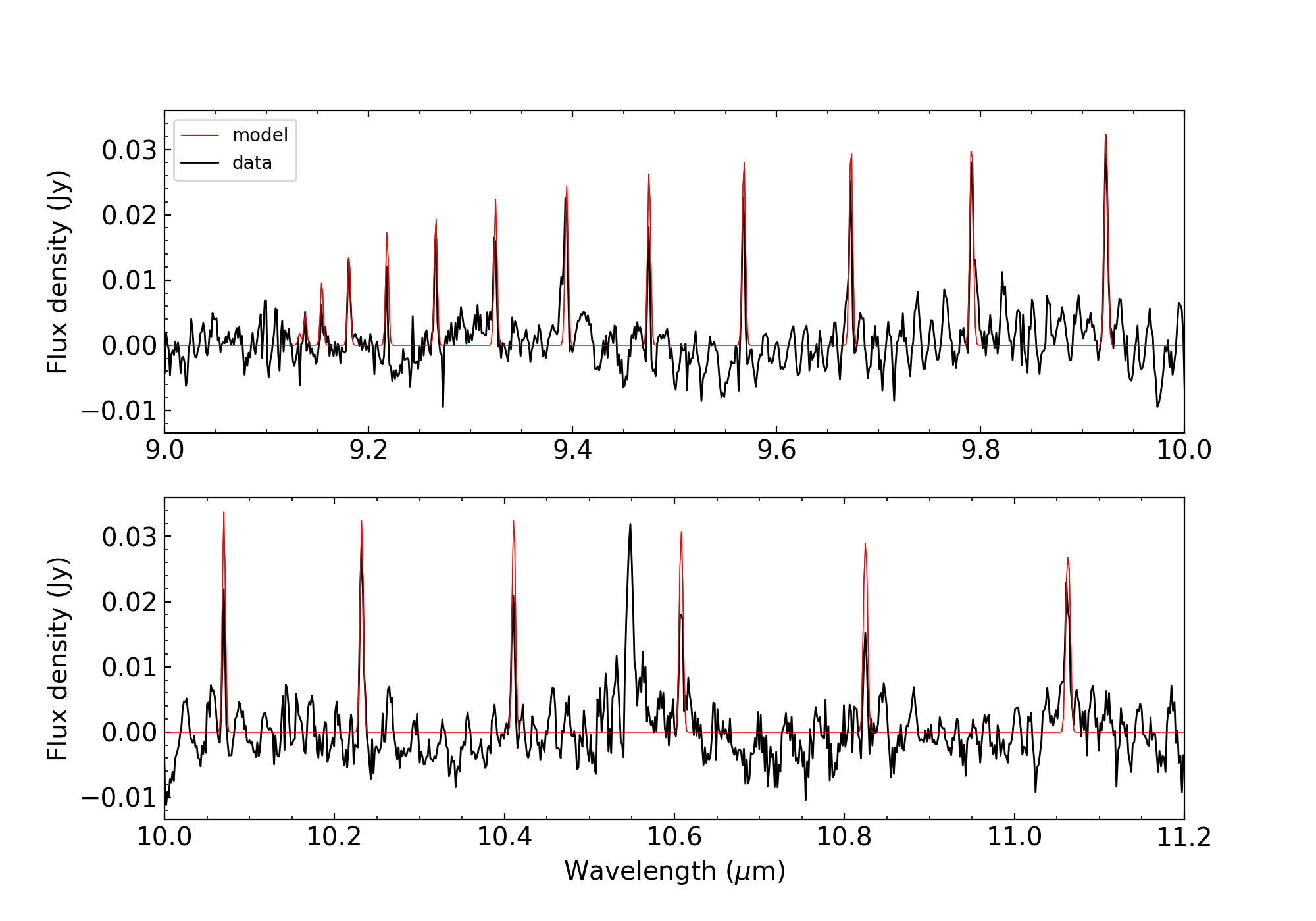}{0.45\textwidth}{(B)}
}
\gridline{\fig{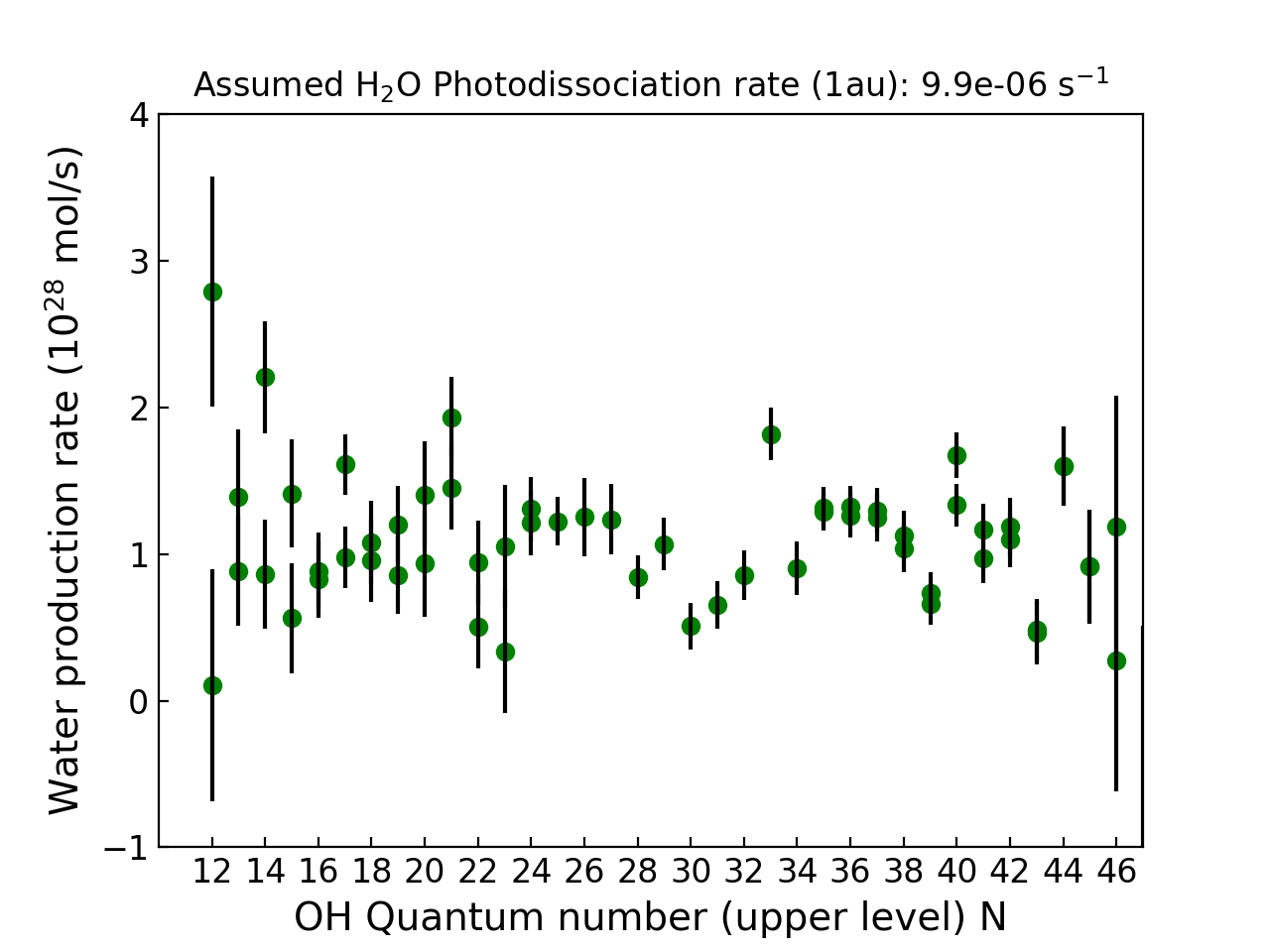}{0.45\textwidth}{(C)}
}
\caption{\textbf{(A).} Full OH$^*$ mid-infrared spectrum of comet E3 (black) with best-fit model overplotted in red. The upper OH$^*$ quantum number, $N$, is noted for each transition. \textbf{(B).} As in Panel A, but zooomed to show the fit in the $9-11$ \um{} region. \textbf{(C).} Derived $Q(\ce{H2O})$ for each OH$^*$ transition as a function of upper level quantum number $N$.
\label{fig:oh-panels}}
\end{figure*}

As noted in \S~\ref{sec:results}, there was strong OH$^*$ emission present in the E3 JWST spectra, including some that were not fit well by the PSG in the G395H setting. In addition to these, we identified multiple OH$^*$ mid-infrared transitions (Figure~\ref{fig:oh-panels}). We applied the emission model of \cite{Tabone2021} and calculated $Q(\ce{H2O})$ for each identified OH$^*$ transition. The results demonstrate an overall good match between the model and observed transitions. Although the $Q(\ce{H2O})$ calculated from individual low-$N$ OH$^*$ lines are consistent with that derived from the simultaneously measured 6 \um{} \ce{H2O} in the same aperture ($Q(\ce{H2O})=(2.56\pm0.16)\times10^{28}$ \ps{}; Table~\ref{tab:qs}), the values are lower for the high-$N$ OH$^*$ lines. This may be related to the assumed \ce{H2O} photodissociation rate, which is the subject of future work.

\subsection{Thermal Modeling of Continuum Emission}\label{subsec:thermal}

Comets contain refractory dust grains that are a mixture of ISM dust and solar nebula processed material.
The grains are composed primarily of silicates (amorphous and crystalline),
that produce distinct emission features in the mid-IR, and dark absorbing material producing
a featureless underlying continuum.  

As a first order look at the refractory dust grains in the coma of 
E3, we modeled the MIRI spectrum using a thermal dust grain model \citep{Harker2023}.
The model uses five distinct material components: amorphous and crystalline olivine and pyroxene for
the silicate material, and amorphous carbon representing the dark absorbing material.  The amorphous
grains have a fractal porosity parameter with a fractional filled volume given by
$f = 1 - (a/0.1~\mu m)^{D-3}$; the fractal dimension parameter $D$ ranging
from $D=3$ (solid) to $D=2.5$ (fractally porous).  Five values of $D$ are chosen to produce the 
thermal model grids (3.0, 2.857, 2.727, 2.609, and 2.5).  The absorbing efficiency of the amorphous
grains are calculated using Mie Theory, with the fractally porous grains generated using
Bruggeman effective medium theory \citep{Bohren1983}.  The crystalline grains have three
crystallographic axes and are not well modeled by Mie Theory; therefore, they are modeled using
a continuous distribution of ellipsoids \citep{Harker2007}, and range in size from 0.1 -- 1~\micron{}
in effective radius.

The differential grain size distribution is represented by the Hanner modified power law \citep[hereafter HGSD;][]{Hanner1994}.  The number of each grain is calculated by 
$n(a) = (1 - a_{\circ}/a)^M (a_{\circ}/a)^N$, where $a$ is the grain radius, $a_{\circ} = 0.1$~\micron, 
the minimum grain radius, and $M$ and $N$ are independent parameters.  $M$ is used with $N$ to calculate 
the peak of the HGSD; $a_{peak} = (M + N)/N$.  A linear combination of each material is least-sqaures
fit to the SED of E3.

\begin{figure}
\plotone{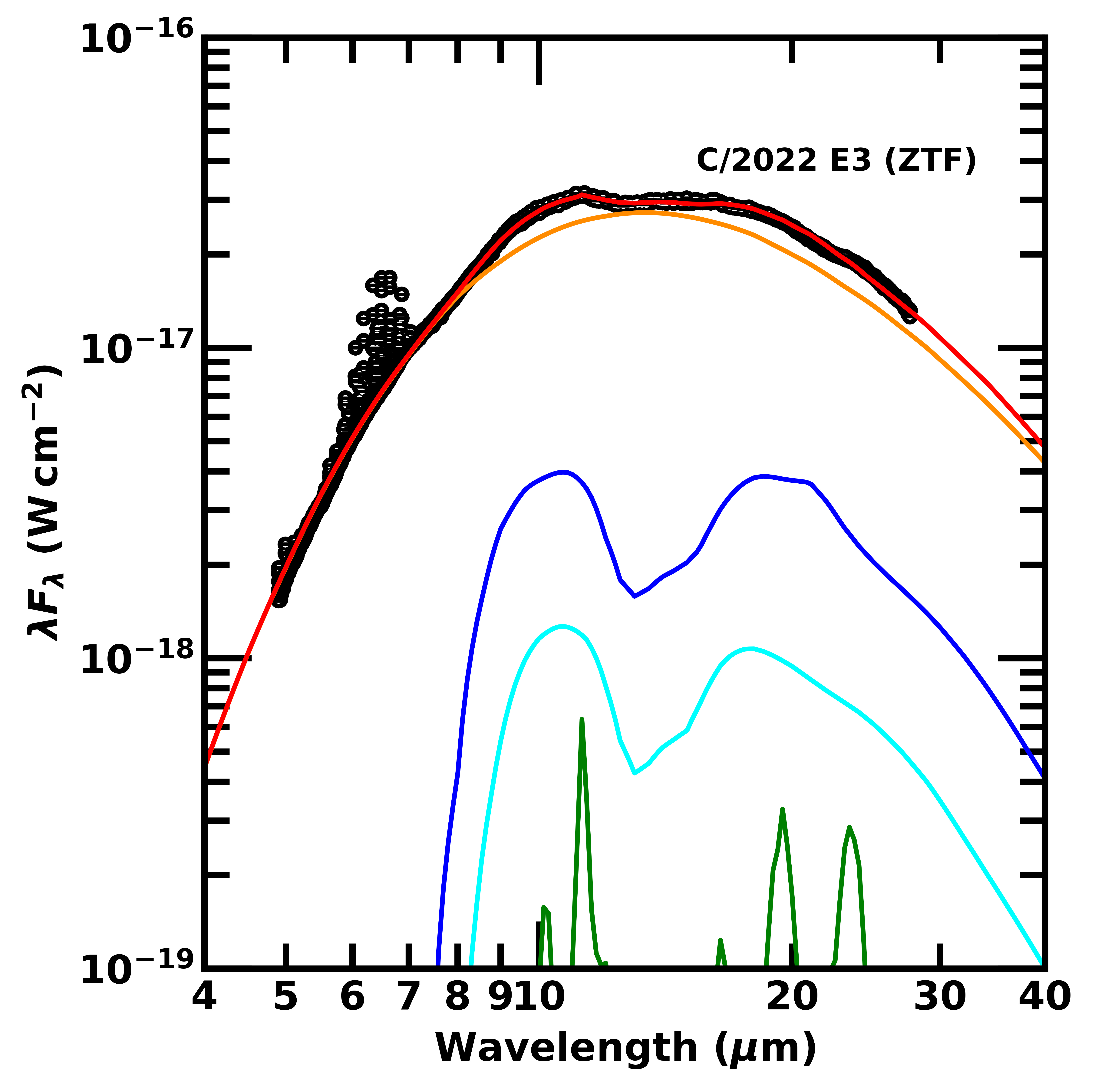}
\caption{Best-fit thermal model (red) for the MIRI MRS SED of C/2022 E3 extracted in a 1\farcs3 diameter nucleus-centered aperture (black). The models provide the dust mineralogy assuming that its constituents have optical constants similar to those of amorphous carbon (orange), amorphous pyroxene (Mg:Fe = 50:50; blue), amorphous olivine (Mg:Fe = 50:50, cyan), Mg-cyrstalline olivine (forsterite; green), and (if present) Mg-crystalline orthopyroxene (enstatite; pink).
\label{fig:e3-model}}
\end{figure}

The MIRI SED of comet E3 covers a wavelength range from $4.9 - 27.9$~\micron.  That range includes
water and gas bands in the $6 - 7$~\micron{} region that are masked out when performing the thermal
fits.  Furthermore, data points longer than 18~\micron{} are also masked out due to uncertainties
in the absolute flux calibration at larger wavelengths.  Outside of these masked points, no other processing of the data (e.g., smoothing)
was performed.  The resulting model fit is shown in Figure~\ref{fig:e3-model}, and the best-fit model parameters are given in Tables~\ref{tab:thermal-particles} and~\ref{tab:thermal-mass}.

\begin{deluxetable*}{cccccccccccc}
\tablenum{3}
\tablecaption{C/2022 E3 Thermal Emission Model Parameters\label{tab:thermal-particles}}
\tablewidth{0pt}
\tablehead{
\colhead{} & \colhead{} & \colhead{} & \colhead{} & \multicolumn{5}{c}{\underbar{$N_p (\times 10^{18})$\sups{b} }} & \colhead{} & \colhead{} \\
\colhead{} & \colhead{} & \colhead{} & \colhead{} & \colhead{Amorphous} & \colhead{Amorphous} & \colhead{Amorphous} & \colhead{Crystalline} & \colhead{Crystalline} & \colhead{} & \colhead{} \\
\colhead{$N$} & \colhead{$M$} & \colhead{$a_p$\sups{a}} & \colhead{$D$} & \colhead{Pyroxene} & \colhead{Olivine} & \colhead{Carbon} & \colhead{Olivine} & \colhead{Orthopyroxene} & \colhead{$\chi^2_\nu$} & \colhead{$\chi^2$}\\
\colhead{} & \colhead{} & \colhead{(\um{})} & \colhead{} & \colhead{(AP50)} & \colhead{(AO50)} & \colhead{(AC)} & \colhead{(CO)} & \colhead{(CP)} & \colhead{} & \colhead{} \\ 
}
\startdata
4.3 & 30.1 & 0.3 & 2.727 & $1.7242^{+0.0020}_{-0.0019}$ & $0.286^{+0.0011}_{-0.0011}$ & $7.44299^{+0.00031}_{-0.00031}$ & $0.356^{+0.0017}_{-0.0018}$ & ... & 245.3 & 2265921 \\
\enddata
\tablecomments{Ellipses indicate the compositional inclusion is not required to achieve the fit. \sups{a} Derived parameter. \sups{b} Number of grains, $N_p$, at the peak ($a_p$) of the Hanner grain size distribution (HGSD).}
\end{deluxetable*}

\begin{deluxetable*}{cccccccc}
\setlength{\tabcolsep}{4pt}
\tablenum{4}
\tablecaption{C/2022 E3 Mass Fractions of Sub-micron Grains\label{tab:thermal-mass}}
\tablewidth{0pc}
\tablehead{
\colhead{Total} & \colhead{Amorphous} & \colhead{Amorphous} & \colhead{Amorphous} & \colhead{Crystalline} & \colhead{Crystalline} & \colhead{Silicate/Carbon} & \colhead{} \\
\colhead{Mass} & \colhead{Pyroxene} & \colhead{Olivine} & \colhead{Carbon} & \colhead{Olivine} & \colhead{Orthopyroxene} & \colhead{Ratio} & \colhead{$f_{\mathrm{cryst}}$} \\
\colhead{$(\times10^7$ kg)} & \colhead{$(f(\mathrm{ap50})\times10^{-1})$} & \colhead{$(f(\mathrm{ao50})\times10^{-1})$} & \colhead{$(f(\mathrm{ac})\times10^{-1})$} & \colhead{$(f(\mathrm{co})\times10^{-1})$} & \colhead{$(f(\mathrm{cp})\times10^{-1})$} & \colhead{$(\times10^{-1})$} & \colhead{$(\times10^{-1})$} \\
}
\startdata
$1.04610^{+0.00057}_{-0.00058}$ & $2.8624^{+0.0028}_{-0.0029}$ & $0.4756^{+0.0019}_{-0.0020}$ & $5.6162^{+0.0031}_{-0.0030}$ & $0.1045^{+0.0045}_{-0.0047}$ & ... & $7.8055^{+0.0097}_{-0.0100}$ & $2.3853^{+0.0087}_{-0.0090}$ \\ 
\enddata
\tablecomments{Asymmetric uncertainties indicate values constrained to a confidence level fo 95\%. Ellipses indicate the compositional inclusion is not required to achieve the fit. Total Mass is derived for emission within a nucleus-centered 1\farcs3 diameter aperture.}
\end{deluxetable*}

\begin{figure}
\plotone{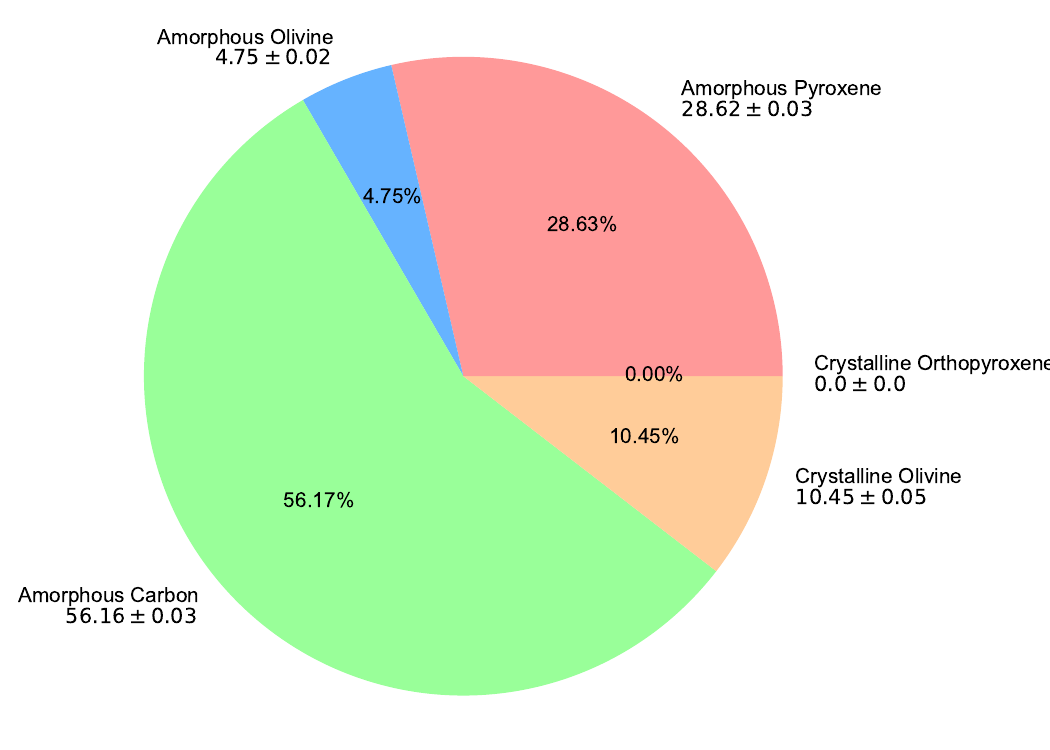}
\caption{Pie diagram of the best-fit thermal mass fractions for constituents of the sub-micron dust grains of C/2022 E3 (Table~\ref{tab:thermal-mass}).
\label{fig:e3-pie}}
\end{figure}

\begin{figure}
\plotone{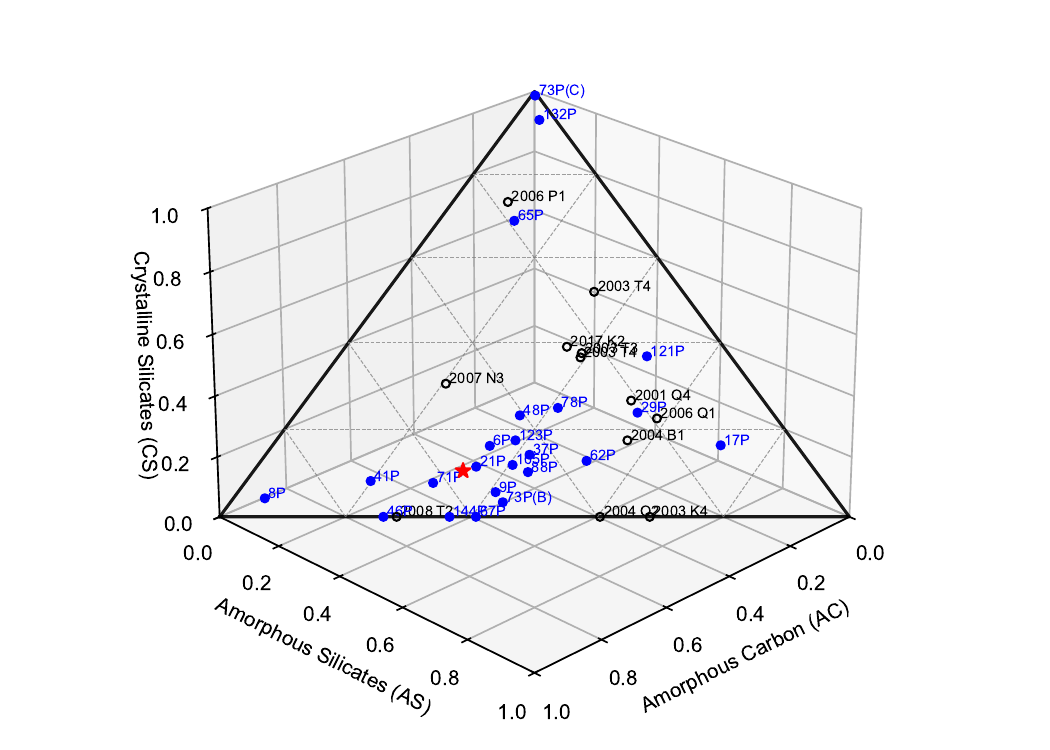}
\caption{Ternary diagram comparing the sub-micron dust grain constituent mass fractions for C/2022 E3 (red star) against comets surveyed by Spitzer \citep{Harker2023} and JWST measurements of C/2017 K2 \citep{Woodward2025}. Oort cloud comets are represented by unfilled black circles, and Jupiter-family comets by filled blue circles. The three components are the sum of the constituent materials (AS = amorphous olivine $+$ amorphous pyroxene, CS = crystalline olivine $+$ crystalline orthopyroxene, AC = amorphous carbon) whose sum is equal to 1. The diagram represents dust compositions which sum to unity in a triangular plane projected into three-dimensional space.
\label{fig:e3-ternary}}
\end{figure}

The materials in the resulting best fit include amorphous pyroxene and olivine, crystalline olivine, and
amorphous carbon.  There was no amount of crystalline pyroxene (orthopyroxene) in the best fit. A diagram of the mass fractions of each component is given in Figure~\ref{fig:e3-pie}, and a ternary diagram comparing the mass fractions for E3 against comets measured in a Spitzer survey of comets (as well as JWST measurements of C/2017 K2) is given in Figure~\ref{fig:e3-ternary}.  The peak of the
HGSD is 0.8~\micron{} and a fractal porosity for the amorphous grains of $D = 2.727$.  The crystalline
mass fraction of the submicron sized silicate, defined as 
$f_{cryst}^{silicates} \equiv$ (crystalline)/(crystalline + amorphous), is $f = 0.239$.  This is slightly lower than
the crystalline silicate mass fraction for C/1995 O1 (Hale-Bopp) and 9P/Tempel~1 post Deep Impact which
are both around 30\% \citep{Woodward2021, Harker2002, Harker2005, Harker2007}. Figure~\ref{fig:e3-ternary} shows that among measured comets, E3 was most similar to Jupiter-family comets 21P/Giacobini-Zinner and 71P/Clark in terms of the relative mass fractions of amorphous and crystalline dust species in its coma.

\section{Conclusion} \label{sec:conclusion}
We have presented spatial-spectral maps of the coma of C/2022 E3 measured with JWST. Our analysis of the molecular emission showed a comet that was largely depleted in its trace volatile abundances compared to mean values in comets, with the exception of \ce{CO2} and \ce{C2H6}. Our spatial-spectral maps showed a complex coma with \ce{CO2}, \ce{^13CO2}, \ce{CH4}, CO, and \ce{CH3OH} having similar spatial distributions of column density, whereas \ce{H2O} was distinct from all of these. However, all molecules showed a consistent \trot{} spatial distribution, with enhanced values in the anti-sunward hemisphere, consistent with previous work analyzing a subset of these data \citep{Foster2026}. Our derived values for the \ce{H2O} OPR did not deviate significantly from the equilibrium value of 3, and the \ce{^12CO2}/\ce{^13CO2} ratio was consistent with the terrestrial value. The dust composition of C/2022 E3 was dominated by amorphous carbon by at least a factor of two relative to other modeled constituents. These results add to the growing body of comets analyzed by the powerful spatial-spectral mapping capabilities of JWST, which provides remarkable insights into the composition and thermal physics of comet comae across the solar system.

\begin{acknowledgments}
This work is based on observations made with the NASA/ESA/CSA James Webb Space Telescope. The data were obtained from the Mikulski Archive for Space Telescopes at the Space Telescope Science Institute, which is operated by the Association of Universities for Research in Astronomy, Inc., under NASA contract NAS 5-03127 for JWST. These observations are associated with program \#1253. The specific observations analyzed can be accessed via doi:10.17909/v0me-1043. N.X.R. is supported by the NASA Planetary Science Division Internal Scientist Funding Program through the Fundamental Laboratory Research work package (FLaRe). H.B.H. and S.N.M. acknowledge support from NASA JWST Interdisciplinary Scientist grant 21-SMDSS21-0013. Part of this research was conducted at the Jet Propulsion Laboratory, California Institute of Technology, under a contract with the National Aeronautics and Space Administration.
\end{acknowledgments}

\software{Astropy \citep{astropy:2013, astropy:2018, astropy:2022},
Astroquery \citep{Ginsburg2019},
photutils \citep{bradley2025},
jwstComet \citep{Roth2026b}
}


\bibliography{E3}{}
\bibliographystyle{aasjournalv7}



\end{document}